# Enhanced MIMO-DCT-OFDM System Using Cosine Domain Equalizer

**Khaled Ramadan**

Department of Communications and Computer Engineering,

The Higher Institute of Engineering at Al-Shorouk City, Cairo, Egypt

Email: ramadank637@gmail.com

*Abstract*—**Discrete Cosine Transform (DCT) can be used instead of conventional Discrete Fourier Transform (DFT) for the Orthogonal Frequency Division Multiplexing (OFDM) construction, which offers many advantages. In this paper, the Multiple-Input-Multiple-Output (MIMO) DCT-OFDM is enhanced using a proposed Cosine Domain Equalizer (CDE) instead of a Frequency Domain Equalizer (FDE). The results are evaluated through the Rayleigh fading channel with Co-Carrier Frequency Offset (Co-CFO) of different MIMO configurations. The average bit error probability and the simulated time of the proposed scheme and the conventional one is compared, which indicates the importance of the proposed scheme. Also, a closed formula for the number of arithmetic operations of the proposed equalizer is developed. The proposed equalizer gives a simulation time reduction of about 81.21%, 83.74% compared to that of the conventional LZF-FDE, and LMMSE-FDE, respectively for the case of 4×4 configuration.**

**Nomenclature**

| | |
|---|---|
| $\mathrm{X}^{i} \in \mathbb{R}^{N\times 1}$ | The transmitted $i^{\text{th}}$ data vector before the Inverse DCT (IDCT) block |
| $\mathrm{x}^{i} \in \mathbb{R}^{N\times 1}$ | The IDCT output of the $i^{\text{th}}$ vector |
| $(\#)^{T}$ | The matrix transpose of #. |
| $\mathbf{C}_N^{-1} \in \mathbb{C}^{N\times N}$ | IDCT matrix |
| $\mathbf{C}_N \in \mathbb{C}^{N\times N}$ | DCT matrix |
| $\mathrm{x}_{\mathrm{CP}}^{i} \in \mathbb{R}^{(N+N_{\mathrm{CP}})\times 1}$ | The $i^{\mathrm{th}}$ modulated vector with Cyclic Prefix (CP) |
| $\mathrm{P}_{\mathrm{CP}^{+}} \in \mathbb{R}^{(N+N_{\mathrm{CP}})\times N}$ | The CP adding matrix |
| $N_{\mathrm{CP}}$ | The CP length |
| $\mathbf{I}_{a\times b}$ | An identity matrix of dimension $a \times b$ |
| $\mathbf{0}_{a\times b}$ | A zero matrix of dimension $a \times b$ |
| $\mathrm{y}^{j} \in \mathbb{C}^{(N+N_{\mathrm{CP}})\times 1}$ | The $j^{\mathrm{th}}$ output vector of the CP removal block |
| $\mathrm{z}^{j} \in \mathbb{C}^{N_{\mathrm{CP}}\times 1}$ | The complex Additive White Gaussian Noise (AWGN) vector |
| $\sigma_n^2$ | The noise power |
| $\psi^{j,i} \in \mathbb{C}^{(N+N_{\mathrm{CP}})\times(N+N_{\mathrm{CP}})}$ | The diagonal co-CFO matrix between the $i^{\mathrm{th}}$ transmitting antenna and $j^{\mathrm{th}}$ receiving antenna |
| $\mathcal{H}^{j,i} \in \mathbb{C}^{(N+N_{\mathrm{CP}})\times(N+N_{\mathrm{CP}})}$ | channel Impulse Response Matrix (IRM) between the $i^{\mathrm{th}}$ transmitting antenna and $j^{\mathrm{th}}$ receiving antenna |
| $\varepsilon_{j,i}$ | he normalized co-CFO matrix between the $i^{\mathrm{th}}$ transmitting antenna and $j^{\mathrm{th}}$ receiving antenna |
| $f$ | The frequency shift |
| $\Delta f$ | The sub-carrier spacing |
| $\mathrm{P}_{\mathrm{CP}^{-}} \in \mathbb{R}^{N\times(N+N_{\mathrm{CP}})}$ | The CP removal matrix |
| $\Lambda^{j,i} \in \mathbb{C}^{(N+N_{\mathrm{CP}})\times(N+N_{\mathrm{CP}})}$ | The hybrid co-CFO plus the Rayleigh fading channel matrix between the $i^{\mathrm{th}}$ transmitting antenna and $j^{\mathrm{th}}$ receiving antenna |
| $\mathbf{\Pi}^{j,i} \in \mathbb{C}^{jN\times iN}$ | Full-matrix representation between the $i^{\mathrm{th}}$ transmitting antenna and $j^{\mathrm{th}}$ receiving antenna |
| $\boldsymbol{\mu}^{j,i} \in \mathbb{C}^{jN\times iN}$ | Banded-matrix of $\mathbf{\Pi}^{j,i}$ matrix between the $i^{\mathrm{th}}$ transmitting antenna and $j^{\mathrm{th}}$ receiving antenna |
| $\mathbf{\Upsilon}_{\mathrm{JLCRLZF-CDE}} \in \mathbb{C}^{jN\times iN}$ | The Joint Low Complexity Regularized Linear Zero Forcing-Cosine Domain Equalizer (JLCRLZF-CDE) solution matrix between the $i^{\mathrm{th}}$ transmitting antenna and $j^{\mathrm{th}}$ receiving antenna |
| $\mathbf{\Upsilon}_{\mathrm{LZF-CDE}} \in \mathbb{C}^{jN\times iN}$ | The LZF-CDE equalizer solution matrix between the $i^{\mathrm{th}}$ transmitting antenna and $j^{\mathrm{th}}$ receiving antenna |
| $\mathbf{\Upsilon}_{\mathrm{LMMSE-CDE}} \in \mathbb{C}^{jN\times iN}$ | The LMMSE-CDE equalizer solution matrix between the $i^{\mathrm{th}}$ transmitting antenna and $j^{\mathrm{th}}$ receiving antenna |
| $\tau$ | The banded-matrix bandwidth |
| $\alpha$ | The regularization parameter of the JLCRLZF-CDE |
| $\sigma_X^2$ | The transmitted signal power |
| $\mathrm{R}_{\mathbf{X}}$ | The AWGN covariance matrix |
| $\mathbb{E}\{\#\}$ | The expectation of # |
| $\sigma_{\mathbf{X}}^{2}$ / $\mathrm{R}_{\mathbf{Z}}$ | The SNR value |
| $\bar{\lambda}_q \in \mathbb{C}^{2^{(\sigma-1)}N\times 2^{(\sigma-1)}N}$, $q \in \{1,2,3,4\}$, $\sigma \in \{1,\ 2,3,\dots\}$ | Output matrix of the $(\boldsymbol{\mu}^{H}\boldsymbol{\mu} + \alpha\, \mathbf{I}_{2^{\sigma}N\times 2^{\sigma}N})$ matrix |
| $\bar{\lambda}_q \in \mathbb{C}^{2^{(\sigma-1)}N\times 2^{(\sigma-1)}N}$, $q \in \{1,2,3,4\}$, $\sigma \in \{1,\ 2,3,\dots\}$ | Output inverted matrix of the $(\boldsymbol{\mu}^{H}\boldsymbol{\mu} + \alpha\, \mathbf{I}_{2^{\sigma}N\times 2^{\sigma}N})^{-1}$ matrix |

## I. Introduction

One of the most essential technologies for spectrum efficiency and Inter-Symbol-Interference (ISI) reduction is the OFDM system [1]. This technology is also applicable to 5G and other wireless communication applications. Furthermore, the traditional OFDM system can be built by utilizing the Inverse Discrete Fourier Transform (IDFT) at the transmitter and the DFT at the receiver [1]. Existing OFDM systems are mostly based on conventional OFDM topologies, with the DFT pair serving as multiplexing and de-multiplexing operations. In reality, the DCT pair modules can be used to create another multicarrier scheme that uses a co-sinusoidal set as an orthogonal basis [2], [3]. This system is referred to as Discrete Cosine Transform based Orthogonal Frequency Division Multiplexing (DCT-OFDM) in this research, whereas the standard OFDM system is referred to as DFT-OFDM.

DCT-OFDM has a number of advantages over DFT, including a higher achievable data rate [4], lower computational complexity [5], and is much more resilient against the Inter Carrier Interference (ICI) effect due to the excellent energy concentration and spectral compaction properties inherited by DCT [6]. In addition, because only half of the minimum sub-carrier spacing is necessary to keep sub-carrier orthogonality [7], this results in that the overall number of sub-carriers in the conventional DFT-OFDM system is halved compared to the DCT-OFDM system. As a result, DCT-OFDM is receiving a lot of interest for future wireless communication [8]. This proposed scheme can be applied across various applications, especially in satellite communication, to alleviate bandwidth limitation [9], [10]

In [4], the authors show that the DCT-OFDM increases the achievable data rate over the DFT-OFDM. In [2], the authors demonstrate the superiority of DCT-OFDM in terms of Bit-Error-Rate (BER) performance over DFT-DCT, in addition to the fact that DCT-OFDM can be implemented with reduced computational complexity [5]. The authors in [11] propose a new multicarrier modulation based on the Type-I even discrete cosine transform, which contains new algorithms for channel estimation and signal reconstruction. Also, authors in [11] show that the numerical simulations demonstrate the proposed approach's good performance in terms of channel estimation and information recovery. The authors in [12], examine combined user activity detection and channel estimation for massive grant-free transmission across frequency selective fading channels using DCT-OFDM. The authors of [13] improved the DCT-OFDM index modulation by introducing supplemental index bit assisted transmit diversity. The simulation results in [13] show that the proposed system outperforms both the traditional EDCT-OFDM with index modulation and the DFT-OFDM with Alamouti coding, even with imperfect channel estimation.

The simple FDE is one of the primary advantages of the DFT-OFDM system. On the other hand, the disadvantage for FDE implementation for DCT-OFDM is the need of an extra DFT/IDFT to accomplish the equalization process in the frequency domain [14], [15]. In this paper, we propose a CDE for DCT-OFDM instead of FDE for DCT-OFDM with lower-complexity using banded-matrix approximation [16]. The proposed equalizer is called JLCRLZF-CDE, which executes the equalization and CFO compensation operations concurrently without the use of additional DFT/IDFT blocks. Moreover, the DCT-OFDM research is still in its infancy, and there is plenty of opportunity for additional performance enhancement. The following is a summary of this paper's contributions:

- For a $2^{\sigma} \times 2^{\sigma}$ MIMO-DCT-OFDM systems, $\sigma = \{1,2,3,4,\dots.\}$, the JLCRLZF-CDE is proposed as a suitable name for a tool that executes the equalization and CFO compensation procedures together.
- The proposed equalizer reduces the need for complex multiplication and additions and multiplication in a number of ways, such as by using the banded-matrix approximation [16], to work out how many points in a row should be ignored.

- A reduction in the simulation time needed for complete transceiver structure of the proposed JLCRLZF-CDE with different schemes is compared.
- A closed formula for the number of arithmetic operations of the proposed equalizer is developed compared to other schemes.
- The proposed approach employs a constant value known as the regularization parameter to reduce the impact of the noise amplification issue.
- Simulation results show that the proposed equalizer is robust against the occurrence of estimation errors, as well as various values of the normalized co-CFO.

The paper is organized as follows: Section II represents a general explanation of the proposed DCT-OFDM system model. The proposed JLCRLZF-CDE is presented in section III. The simulation results and analysis is given in section IV. A detailed complexity analysis and closed formula of the proposed JLCRLZF-CDE is presented, then a comparison with respect to conventional schemes is discussed in section V. Finally, the concluded marks are given in section VI.

## II. Proposed DCT-OFDM System Model

The most significant restriction of DCT-OFDM is that the circular convolution property, which is always met by DFT, does not apply to DCT [3], [17]. To solve this issue and enhance the performance of DCT-OFDM even more, the proposed JLCRLZF-CDE is introduced. Figure 1 shows the transmitter and receiver structures of $i\times j$ MIMO-DCT-OFDM system.

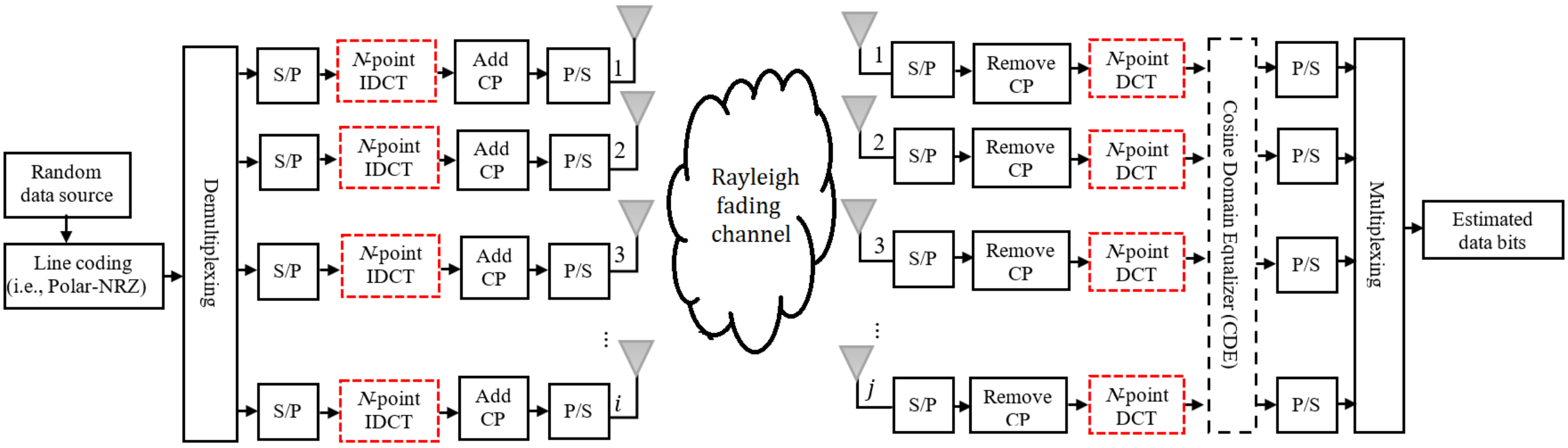

**Figure 1**. $i\times j$ MIMO-OFDM structure based on DCT using CDE.

The produced bits from the random data source are represented by polar NRZ, which is then de-multiplexed through each stream based on the number of transmitting antennas. Each group of $N$ length is passed through a Serial-to-Parallel (S/P) converter. This is followed by IDCT. Thus, the transmitted data vector before the IDCT block associated to the $i^{\text{th}}$ vector can be expressed as:

$$\mathrm{X}^i = [X_0^i \quad X_1^i \quad \ldots\ldots \quad X_{N-2}^i \quad X_{N-1}^i]^T \tag{1}$$

where $\mathrm{X}^i \in \mathbb{R}^{N\times 1}$, $N$ is the sub-carrier number, and $(\#)^T$ is the matrix transpose of #. Now, the IDCT block is applied to each $N$ vector. As a result, the IDCT output of the $i^{\text{th}}$ vector can be expressed as follows:

$$\mathrm{x}^i = \mathbf{C}_N^{-1}\,\mathrm{X}^i = [x_0^i \quad x_1^i \quad \ldots\ldots \quad x_{N-2}^i \quad x_{N-1}^i]^T \tag{2}$$

where $\mathrm{x}^i \in \mathbb{R}^{N\times 1}$, $\mathbf{C}_N^{-1}$ is an $N$-point IDCT, whose elements are defined as [18]:

$$[\mathbf{C}_N^{-1}]_{k,n} = \beta_n \cos\left(\frac{\pi}{2N}(2k-1)(n-1)\right), k,n = 1,2,..,N \tag{3}$$

with

$$k,n \in \{1,2,\ldots N\}, \qquad \beta_n = \begin{cases} \frac{1}{\sqrt{N}} & n = 1 \\ \sqrt{\frac{2}{N}} & n \neq 1 \end{cases} \tag{4}$$

Now, the CP is added to the head of each vector. Hence, the $i^{\text{th}}$ modulated vector with CP can be formulated as:

$$\mathrm{x}_{\mathrm{CP}}^i = \mathrm{P}_{\mathrm{CP}}\,\mathrm{x}^i = \left[x_{0,\mathrm{CP}}^i \quad x_{1,\mathrm{CP}}^i \quad \ldots\ldots \quad x_{N-2,\mathrm{CP}}^i \quad x_{N-1,\mathrm{CP}}^i\right]^T \tag{5}$$

with

$$\mathrm{P}_{\mathrm{CP}^+} = \left[[\mathbf{0}_{N_{\mathrm{CP}}\times(N-N_{\mathrm{CP}})}; \mathbf{I}_{N_{\mathrm{CP}}\times N_{\mathrm{CP}}}]^T, \mathbf{I}_{N\times N}\right]^T \tag{6}$$

where $\mathrm{P}_{\mathrm{CP}^+} \in \mathbb{R}^{(N+N_{\mathrm{CP}})\times N}$, $\mathrm{x}_{\mathrm{CP}}^i \in \mathbb{R}^{(N+N_{\mathrm{CP}})\times 1}$, $N_{\mathrm{CP}}$ denotes the CP length, $\mathbf{I}_{a\times b}$ is an identity matrix of dimension $a \times b$, and $\mathbf{0}_{a\times b}$ is a zero matrix of dimension $a \times b$. Now, the Parallel-to-Serial (P/S) converter is applied for each group then transmitted over the Rayleigh fading channel.

At the receiver, the received vector at the $j^{\text{th}}$ antenna, which includes the Rayleigh fading channel, co-CFO effect, as well as the noise after the S/P converter, is given by:

$$\mathrm{y}^{j} = \psi^{j,i}\, \mathcal{H}^{j,i} \mathrm{P}_{\mathrm{CP}^{+}}\, \mathrm{x}^{i} + \mathrm{z}^{j} \tag{7}$$

where $\mathrm{y}^{j} \in \mathbb{C}^{(N+N_{\mathrm{CP}})\times 1}$ represents the $j^{\text{th}}$ output vector of the CP removal block, $\mathrm{z}^{j} \in \mathbb{C}^{N_{\mathrm{CP}}\times 1}$ denotes the complex AWGN vector, complex Gaussian $\backsim \mathcal{CN}(0,\sigma_n^2)$ and $\sigma_n^2$ is the noise power, $\psi^{j,i} \in \mathbb{C}^{(N+N_{\mathrm{CP}})\times(N+N_{\mathrm{CP}})}$ is a diagonal matrix, which represents the co-CFO matrix between the $i^{\text{th}}$ transmitting antenna and $j^{\text{th}}$ receiving antenna, which is defined as:

$$\psi^{j,i} = \mathrm{Diag}\left\{1, e^{j\frac{2\pi\varepsilon_{j,i}}{N}}, e^{j\frac{4\pi\varepsilon_{j,i}}{N}}, \dots, e^{j\frac{2\pi\varepsilon_{j,i}(N+N_{\mathrm{CP}}-2)}{N}}, e^{j\frac{2\pi\varepsilon_{j,i}\,(N+N_{\mathrm{CP}}-1)}{N}}\right\} \tag{8}$$

where $\varepsilon_{j,i}$ represents the normalized co-CFO between the $i^{\text{th}}$ transmitting antenna and $j^{\text{th}}$ receiving antenna, which is defined as:

$$\varepsilon_{j,i} = \frac{f}{\Delta f} \tag{9}$$

where $f$ denotes the frequency shift and $\Delta f$ denotes the sub-carrier spacing. Furthermore, the normalized co-CFO value is affected by a number of parameters, including transform size (i.e. $N$), carrier frequency, oscillator stability, mobile speeds, and signal bandwidth.

$\mathcal{H}^{j,i} \in \mathbb{C}^{(N+N_{\mathrm{CP}})\times(N+N_{\mathrm{CP}})}$ is the channel IRM between the $i^{\text{th}}$ transmitting antenna and $j^{\text{th}}$ receiving antenna, which is defined as:

$$\mathcal{H}^{i,j} = \begin{pmatrix} h(0) & 0 & 0 & 0 & \dots & 0 & 0 & 0 \\ h(1) & h(0) & 0 & 0 & \dots & 0 & 0 & 0 \\ \vdots & h(1) & h(0) & 0 & \dots & 0 & 0 & 0 \\ h(L-1) & \vdots & h(1) & h(0) & \vdots & 0 & 0 & 0 \\ 0 & h(L-1) & \vdots & h(1) & \ddots & 0 & 0 & 0 \\ 0 & 0 & h(L-1) & \vdots & \vdots & h(0) & 0 & 0 \\ 0 & 0 & 0 & h(L-1) & \dots & h(1) & h(0) & 0 \\ 0 & 0 & 0 & 0 & \dots & h(2) & h(1) & h(0) \end{pmatrix} \tag{10}$$

where $h(w)$, is the channel impulse response coefficient, $w \in \{1, 2, \dots, L-1\}$, and $L$ is the number of the Rayleigh fading channel taps. Now, the head of each vector that represents the CP is disregarded. Furthermore, the $j^{\text{th}}$ vector's CP removal block output is given by:

$$\bar{y}^j = \mathrm{P}_{\mathrm{CP}^-}\,\mathrm{y}^j = \mathrm{P}_{\mathrm{CP}^-}\Lambda^{j,i}\mathrm{P}_{\mathrm{CP}^+}\,\mathrm{x}^i + \mathrm{P}_{\mathrm{CP}^-}\mathrm{z}^j \tag{11}$$

where $\Lambda^{j,i} = \psi^{j,i}\,\mathcal{H}^{j,i}$ is the hybrid co-CFO plus the Rayleigh fading channel, $\mathrm{P}_{\mathrm{CP}^-} \in \mathbb{R}^{N\times(N+N_{\mathrm{CP}})}$ represents the CP removal matrix. Using Eq. (2), the previous equation can be represented by:

$$\begin{bmatrix}\bar{y}^1\\ \bar{y}^2\\ \vdots\\ \bar{y}^j\end{bmatrix} = \begin{bmatrix}\mathrm{P}_{\mathrm{CP}^-}\Lambda^{1,1}\mathrm{P}_{\mathrm{CP}^+}\mathbf{C}_N^{-1} & \mathrm{P}_{\mathrm{CP}^-}\Lambda^{1,2}\mathrm{P}_{\mathrm{CP}^+}\mathbf{C}_N^{-1} & \dots & \mathrm{P}_{\mathrm{CP}^-}\Lambda^{1,i}\mathrm{P}_{\mathrm{CP}^+}\mathbf{C}_N^{-1}\\ \mathrm{P}_{\mathrm{CP}^-}\Lambda^{2,1}\mathrm{P}_{\mathrm{CP}^+}\mathbf{C}_N^{-1} & \mathrm{P}_{\mathrm{CP}^-}\Lambda^{2,2}\mathrm{P}_{\mathrm{CP}^+}\mathbf{C}_N^{-1} & \dots & \mathrm{P}_{\mathrm{CP}^-}\Lambda^{2,i}\mathrm{P}_{\mathrm{CP}^+}\mathbf{C}_N^{-1}\\ \vdots & \vdots & \ddots & \vdots\\ \mathrm{P}_{\mathrm{CP}^-}\Lambda^{j,1}\mathrm{P}_{\mathrm{CP}^+}\mathbf{C}_N^{-1} & \mathrm{P}_{\mathrm{CP}^-}\Lambda^{j,2}\mathrm{P}_{\mathrm{CP}^+}\mathbf{C}_N^{-1} & \dots & \mathrm{P}_{\mathrm{CP}^-}\Lambda^{j,i}\mathrm{P}_{\mathrm{CP}^+}\mathbf{C}_N^{-1}\end{bmatrix}\begin{bmatrix}\mathrm{X}^1\\ \mathrm{X}^2\\ \vdots\\ \mathrm{X}^i\end{bmatrix} + \begin{bmatrix}\mathrm{P}_{\mathrm{CP}^-}\mathrm{z}^1\\ \mathrm{P}_{\mathrm{CP}^-}\mathrm{z}^2\\ \vdots\\ \mathrm{P}_{\mathrm{CP}^-}\mathrm{z}^j\end{bmatrix} \tag{12}$$

Now, an $N$-point DCT is applied, whose elements are defined as [18]:

$$[\mathbf{C}_N]_{k,n} = \beta_n \cos\left(\frac{\pi}{2N}(2k-1)(n-1)\right), k,n = 1,2,..,N \tag{13}$$

with

$$k,n \in \{1,2,\dots N\},\qquad \beta_n = \begin{cases}\dfrac{1}{\sqrt{N}} & n = 1\\ \sqrt{\dfrac{2}{N}} & n \neq 1\end{cases} \tag{14}$$

As a result, Eq. (13) will be applied to Eq. (12) to obtain:

$$\begin{bmatrix}\bar{y}^1\\ \bar{y}^2\\ \vdots\\ \bar{y}^j\end{bmatrix} = \begin{bmatrix}\mathbf{C}_N\mathrm{P}_{\mathrm{CP}^-}\Lambda^{1,1}\mathrm{P}_{\mathrm{CP}^+}\mathbf{C}_N^{-1} & \mathbf{C}_N\mathrm{P}_{\mathrm{CP}^-}\Lambda^{1,2}\mathrm{P}_{\mathrm{CP}^+}\mathbf{C}_N^{-1} & \dots & \mathbf{C}_N\mathrm{P}_{\mathrm{CP}^-}\Lambda^{1,i}\mathrm{P}_{\mathrm{CP}^+}\mathbf{C}_N^{-1}\\ \mathbf{C}_N\mathrm{P}_{\mathrm{CP}^-}\Lambda^{2,1}\mathrm{P}_{\mathrm{CP}^+}\mathbf{C}_N^{-1} & \mathbf{C}_N\mathrm{P}_{\mathrm{CP}^-}\Lambda^{2,2}\mathrm{P}_{\mathrm{CP}^+}\mathbf{C}_N^{-1} & \dots & \mathbf{C}_N\mathrm{P}_{\mathrm{CP}^-}\Lambda^{2,i}\mathrm{P}_{\mathrm{CP}^+}\mathbf{C}_N^{-1}\\ \vdots & \vdots & \ddots & \vdots\\ \mathbf{C}_N\mathrm{P}_{\mathrm{CP}^-}\Lambda^{j,1}\mathrm{P}_{\mathrm{CP}^+}\mathbf{C}_N^{-1} & \mathbf{C}_N\mathrm{P}_{\mathrm{CP}^-}\Lambda^{j,2}\mathrm{P}_{\mathrm{CP}^+}\mathbf{C}_N^{-1} & \dots & \mathbf{C}_N\mathrm{P}_{\mathrm{CP}^-}\Lambda^{j,i}\mathrm{P}_{\mathrm{CP}^+}\mathbf{C}_N^{-1}\end{bmatrix}\begin{bmatrix}\mathrm{X}^1\\ \mathrm{X}^2\\ \vdots\\ \mathrm{X}^i\end{bmatrix} + \begin{bmatrix}\mathbf{C}_N\mathrm{P}_{\mathrm{CP}^-}\mathrm{z}^1\\ \mathbf{C}_N\mathrm{P}_{\mathrm{CP}^-}\mathrm{z}^2\\ \vdots\\ \mathbf{C}_N\mathrm{P}_{\mathrm{CP}^-}\mathrm{z}^j\end{bmatrix} \tag{15}$$

## III. Proposed JLCRLZF-CDE

In this section, the proposed JLCRLZF-CDE will be presented. Let's define $\mathbf{\Pi}^{j,i} \in \mathbb{C}^{jN\times iN}$, $\mathbf{\Pi}^{j,i} = \mathbf{C}_N\mathrm{P}_{\mathrm{CP}^-}\Lambda^{j,i}\mathrm{P}_{\mathrm{CP}^+}\mathbf{C}_N^{-1}$. Now, Eq. (15) will be re-expressed as:

$$\begin{bmatrix}\bar{y}^1\\ \bar{y}^2\\ \vdots\\ \bar{y}^j\end{bmatrix} = \begin{bmatrix}\mathbf{\Pi}^{1,1} & \mathbf{\Pi}^{1,2} & \dots & \mathbf{\Pi}^{1,i}\\ \mathbf{\Pi}^{2,1} & \mathbf{\Pi}^{2,2} & \dots & \mathbf{\Pi}^{2,i}\\ \vdots & \vdots & \ddots & \vdots\\ \mathbf{\Pi}^{j,1} & \mathbf{\Pi}^{j,2} & \dots & \mathbf{\Pi}^{j,i}\end{bmatrix}\begin{bmatrix}\mathrm{X}^1\\ \mathrm{X}^2\\ \vdots\\ \mathrm{X}^i\end{bmatrix} + \begin{bmatrix}\rho^1\\ \rho^2\\ \vdots\\ \rho^j\end{bmatrix} \tag{16}$$

where,

$$\rho^j = \mathbf{C}_N\mathrm{P}_{\mathrm{CP}^-}\mathrm{z}^j \tag{17}$$

Figure 2 shows the magnitude of row number 25 of the $\mathbf{\Pi}^{j,i}$ matrices, $j,i = \{1,2\}$ for the case of 2×2 MIMO configuration. The desired sub-carrier is represented by index 25, whereas the remaining sub-carriers indicate interference. Each normalized co-CFO is a random variable with a uniform distribution in [-$\varepsilon_{max}$, +$\varepsilon_{max}$], where $\varepsilon_{max}$ is the maximum normalized co-CFO. The magnitude of the interference generated by any sub-carrier on the 25th sub-carrier diminishes as the distance between these sub-carriers grows, as seen in Fig. 2. As a result, a design parameter τ can be introduced as a threshold level, which specifies the number of sub-carriers that will be taken into account as:

$$\left(\boldsymbol{\mu}^{j,i}\right)_{m,\bar{m}} = \begin{cases} \left(\mathbf{\Pi}^{j,i}\right)_{m,\bar{m}} & |m-\bar{m}| \le \tau \\ 0 & |m-\bar{m}| > \tau \end{cases} \tag{18}$$

where $m-\bar{m}$ is the relative sub-carrier distance, and $\boldsymbol{\mu}^{j,i}$ is the banded-matrix approximation of the $\mathbf{\Pi}^{j,i}$ matrix.

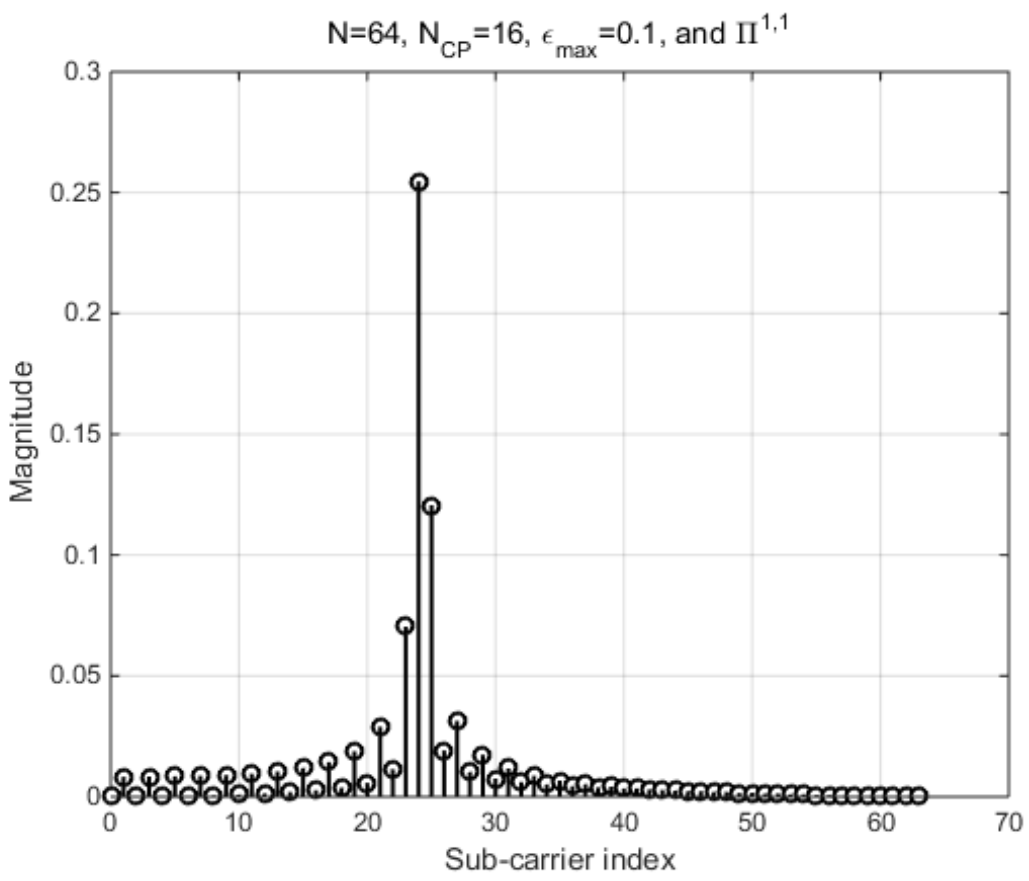

**Figure 2a**. The magnitude versus the sub-carrier index of the $\mathbf{\Pi}^{1,1}$ matrix.

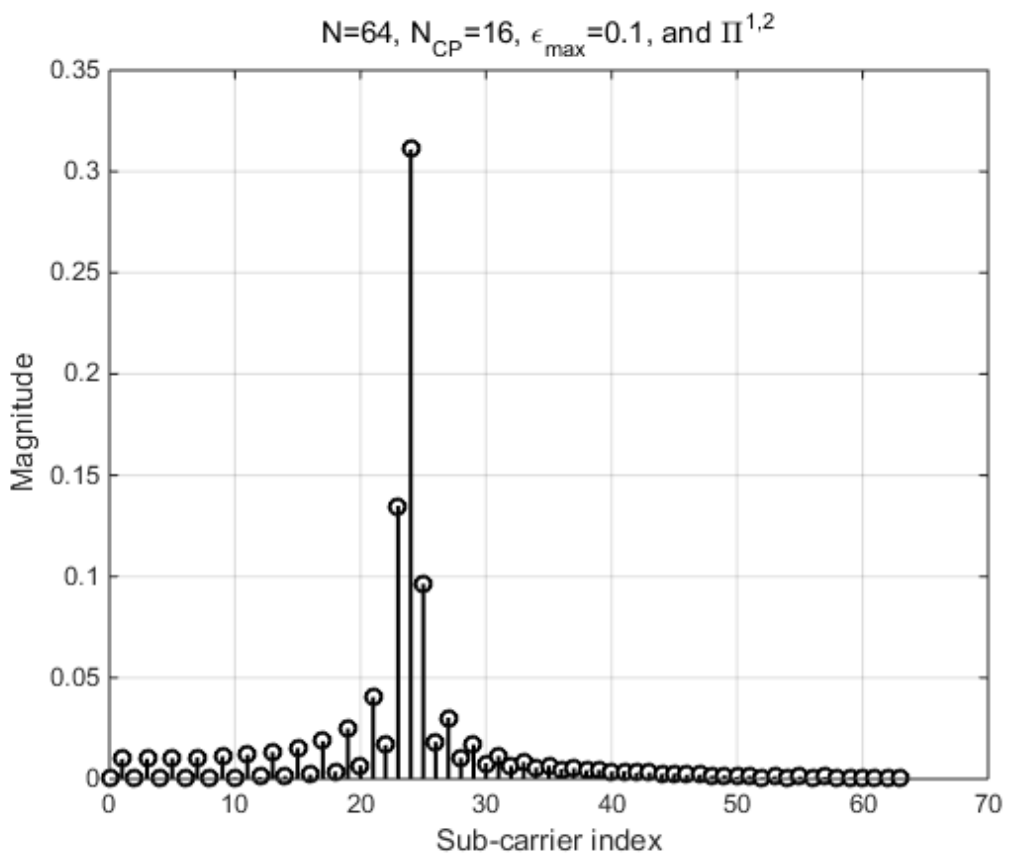

**Figure 2b**. The magnitude versus the sub-carrier index of the $\mathbf{\Pi}^{1,2}$ matrix.

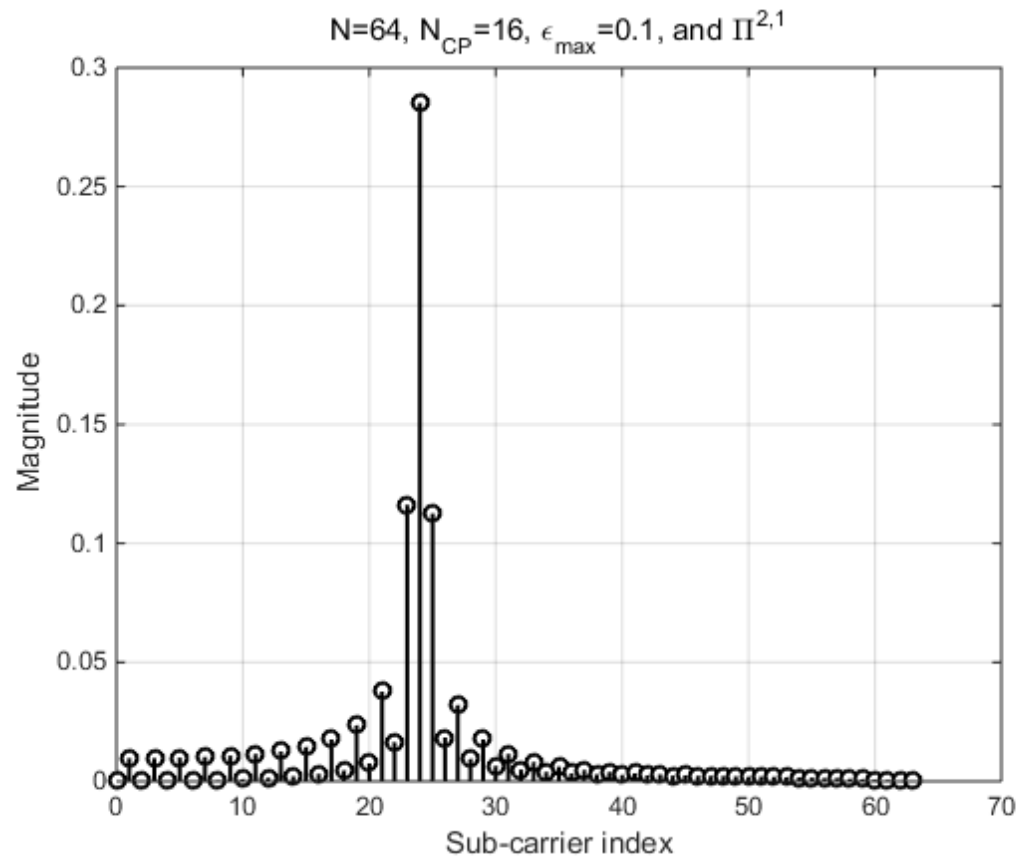

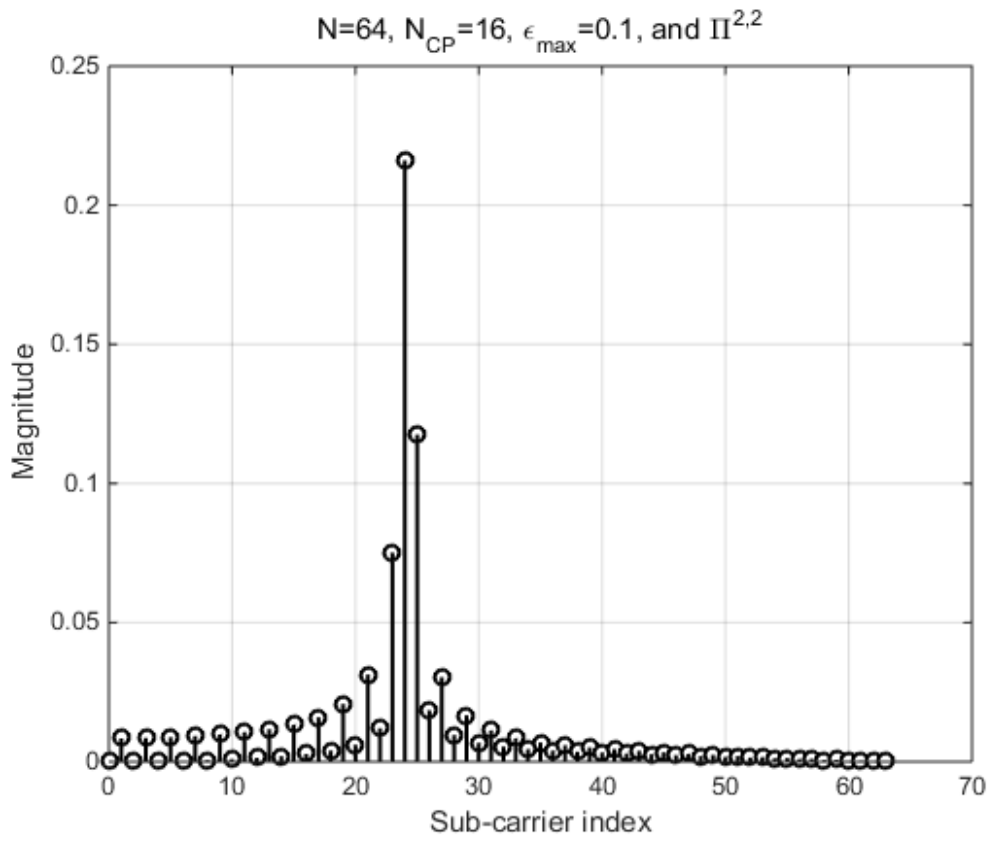

**Figure 2c**. The magnitude versus the sub-carrier index of the $\mathbf{\Pi}^{2,1}$ matrix.

**Figure 2d**. The magnitude versus the sub-carrier index of the $\mathbf{\Pi}^{2,2}$ matrix.

The general matrix solution of the proposed JLCRLZF-CDE is:

$$\mathbf{Y}_{\text{JLCRLZF-CDE}} = \left(\boldsymbol{\mu}^{H}\boldsymbol{\mu} + \alpha\, \mathbf{I}_{2N\times 2N}\right)^{-1} \boldsymbol{\mu}^{H} \tag{19}$$

where α is the regularization parameter that is used for noise enhancement mitigation, $\boldsymbol{\mu} \in \mathbb{C}^{jN\times iN}$ is the banded approximated matrix, that is defined as:

$$\boldsymbol{\mu} = \begin{bmatrix} \boldsymbol{\mu}^{1,1} & \boldsymbol{\mu}^{1,2} & \dots & \boldsymbol{\mu}^{1,i} \\ \boldsymbol{\mu}^{2,1} & \boldsymbol{\mu}^{2,2} & \dots & \boldsymbol{\mu}^{2,i} \\ \vdots & \vdots & \ddots & \vdots \\ \boldsymbol{\mu}^{j,1} & \boldsymbol{\mu}^{j,2} & \dots & \boldsymbol{\mu}^{j,i} \end{bmatrix} \tag{20}$$

As a result, the suggested JLCRLZF-CDE requires the value of the banded-matrix bandwidth (τ) as well as the value of the regularization parameter (α). As a result, the value of the banded-matrix bandwidth (τ) should be as minimal as possible while preserving BER performance. The regularization parameter (α) should be chosen optimally such that the BER performance closely matches that of the Linear Minimum Mean Square Error (LMMSE) equalizer. Note that α=0 in the case of the Linear Zero Forcing (LZF) equalizer, while in the case of LMMSE the value of the Signal-to-Noise Ratio (SNR) must be estimated correctly and α=1/SNR.

## IV. Simulation Results and Analysis

In this section, we present our simulation results for the proposed JLCRLZF-CDE for DCT-OFDM system. According to the simulation parameters listed in Table 1, let's consider a 6-tap Rayleigh fading channel based on the Jak's model [19], with vehicular A model [20]. According to the analysis discussed in the previous section, the values of α and τ should be specified correctly.

**Table 1**. The simulation parameters

| Parameter | Value | Parameter | value |
|---|---|---|---|
| IDCT/DCT size | 64 | SNR range | 0:5:25 |
| IDFT/DFT size | 64 | No. of transmitting antennas | $2^{\sigma}$ |
| CP length | 16 | No. of receiving antennas | $2^{\sigma}$ |
| Channel model | Six-tap Rayleigh fading channel based on Jake's model [19] | Length of transmitted data vector without the CP | 64 |
| Channel type | Frequency-selective fading channel | Simulation type | Monte Carlo |
| $\varepsilon_{max}$ | 0.1 | No. of iterations | $10^3$ |
| Data type | Polar Non-Return to Zero (NRZ) | Noise type | AWGN |

Figure 3a depicts the BER performance versus the change of the regularization parameter (α) at various SNR values. Figure 3b shows the elevation view of Fig. 3a for the case of $\varepsilon_{max}$=0.1. The last two vector values of α=0 and 1/SNR correspond to the LZF and LMMSE equalizer, respectively. Moreover, Fig. 3b shows that the minimum BER performance is achieved at $\alpha=10^{-2}$, and $10^{-3}$. As a result, the BER performance versus SNR should be estimated at $\alpha=10^{-2}$, and $10^{-3}$ to specify the value of α, which used for the rest of simulations.

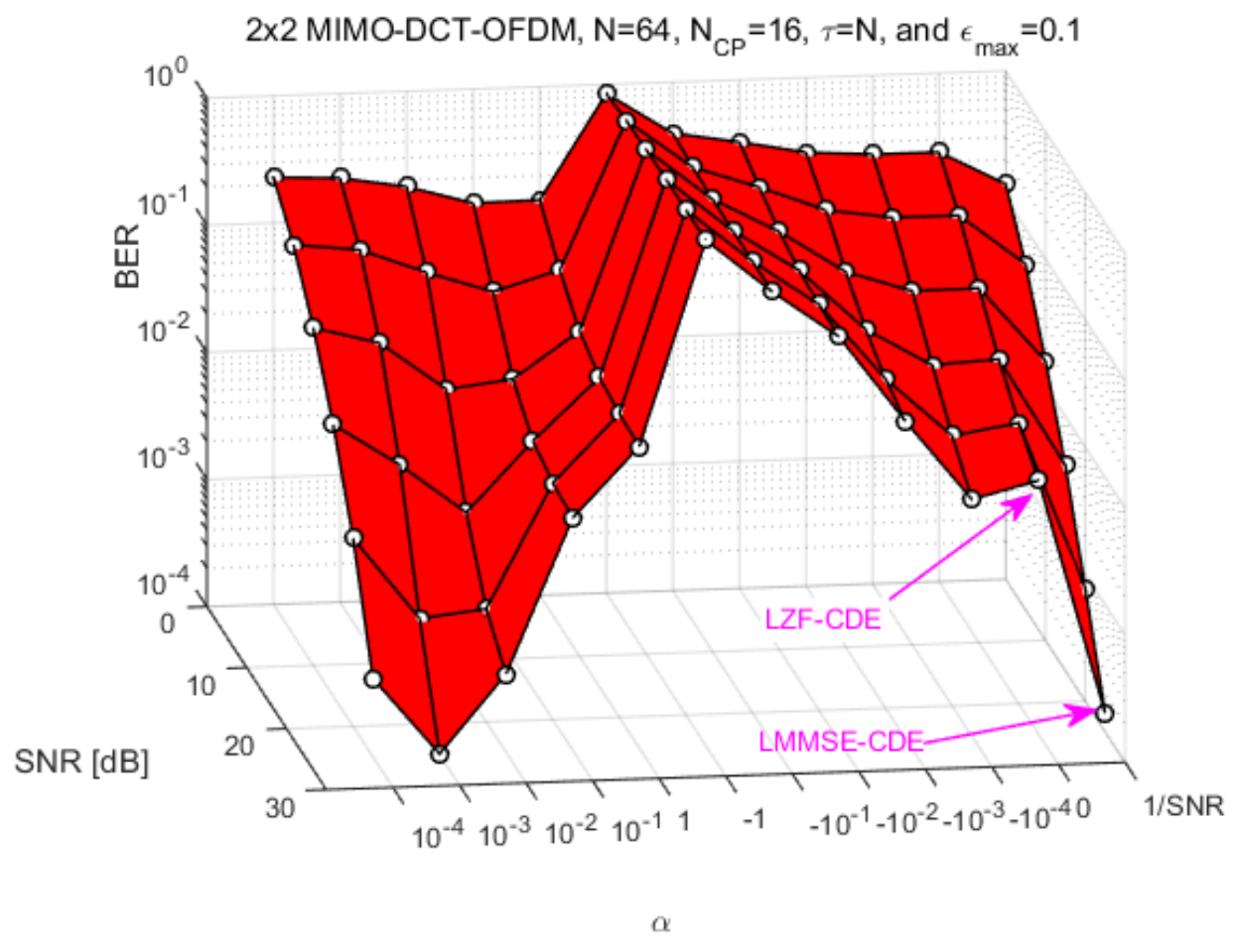

**Figure 3a**. The BER versus the regularization parameter (α) at different values of the SNR.

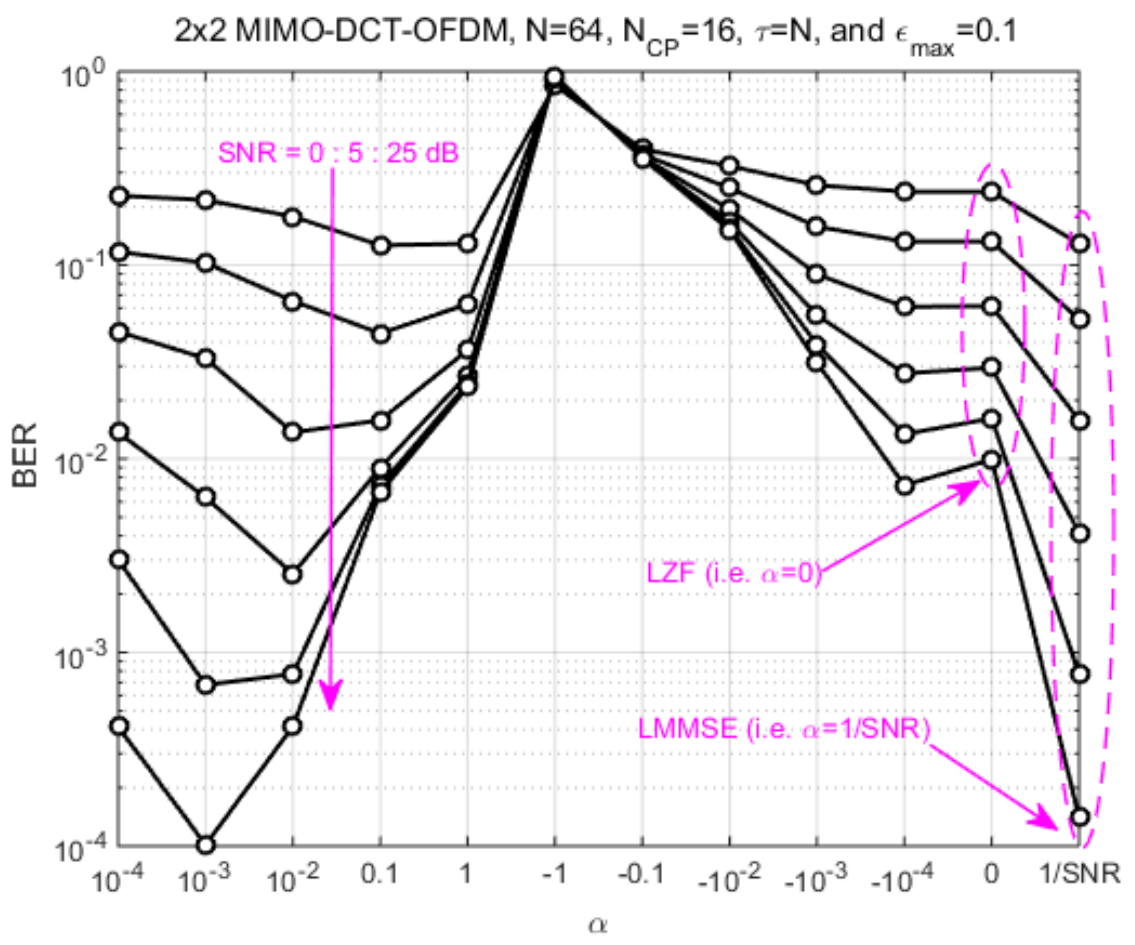

**Figure 3b**. The elevation view of Fig. 3a at $\varepsilon_{max}$=0.1.

Figure 4 depicts the performance of the BER versus the SNR at various levels of the regularization parameter (α). Over the variation range of the SNR, it is evident that $\alpha=10^{-2}$ is the optimum choice, which is matched with the LMMSE equalization. As a result, for the remainder of the simulations, use α =$10^{-2}$. We are now interested in determining the banded-matrix bandwidth (τ) in order to achieve equalization with a reduced complexity design. Figure 5a depicts the BER performance versus the banded-matrix bandwidth (τ) at various SNR levels. The elevation view of Fig. 5a is shown in Figure 5b. The last two vector values in Fig. 5b correspond to the full compensation situation (i.e., τ=*N*) and the no CFO scenario (i.e., ε=0). It is obvious that the BER is primarily saturated as the banded-matrix bandwidth ≥15. As a result, we'll use τ=15 for the rest of the simulations. Figure 6 depicts the BER performance versus the SNR in various compensating situations. Cases with full compensation (τ=*N*) and no CFO (ε=0) are generally matched over the SNR variation range (0≤SNR≤15 dB). On the other hand, the BER performance degrades when the banded-matrix bandwidth decreases. As a result, there is a trade-off between BER performance and the computing complexity of the equalization procedure. At BER=$10^{-3}$ and τ=5, 10, and 15, the SNR degrades by about 6.08, 4.83, and 2.63 dB, respectively, as compared to the full compensation performance scenario. Thus, for the next simulations, let's choose τ=15.

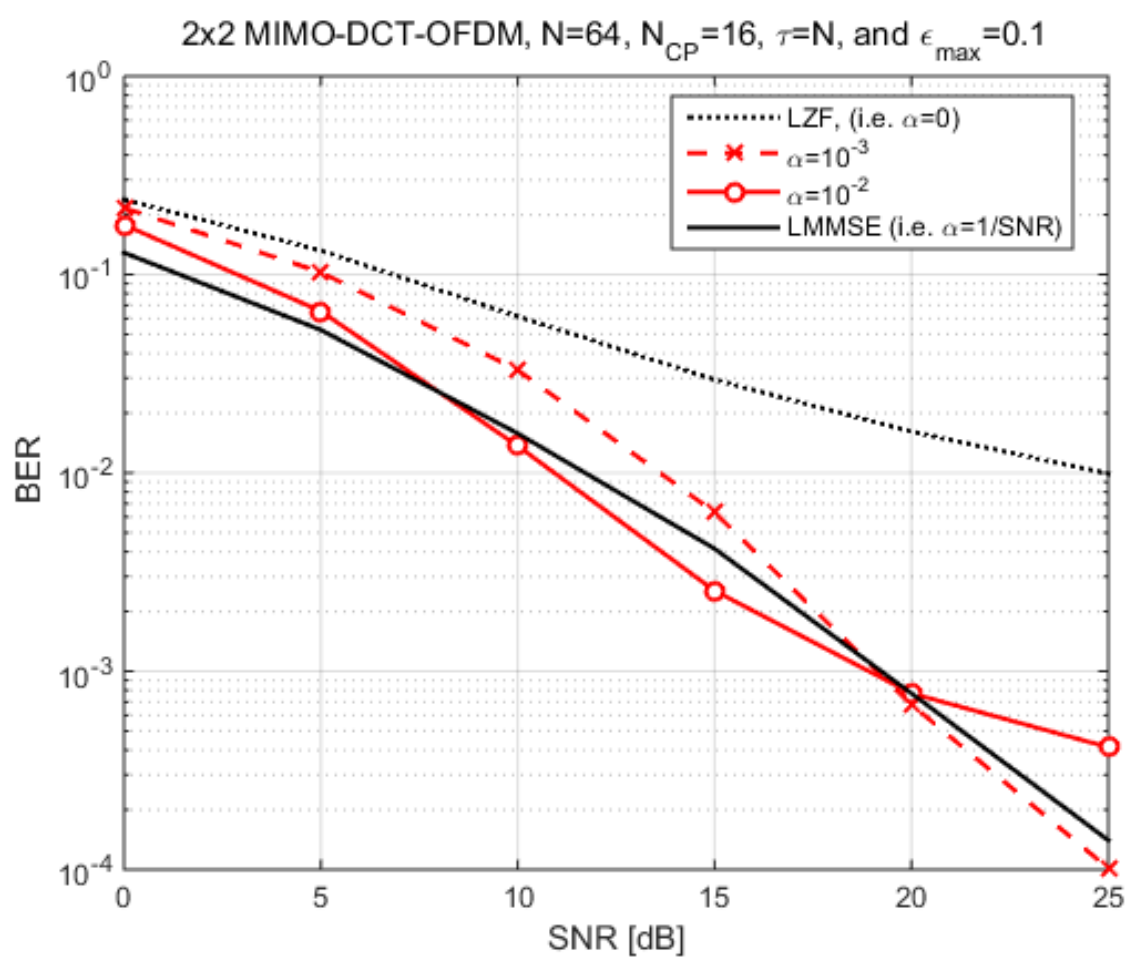

**Figure 4**. The BER versus the SNR at different values of the regularization parameter (α)

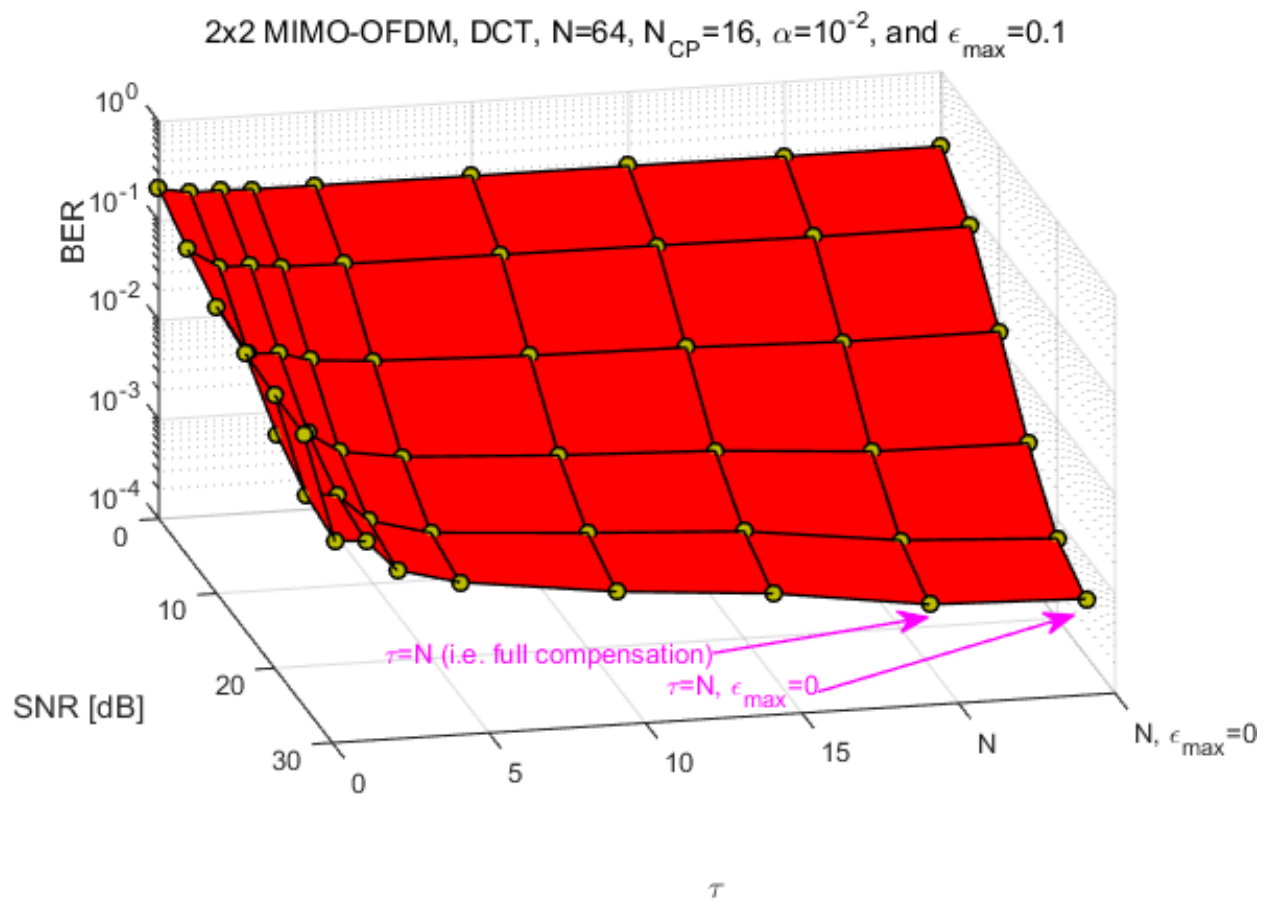

**Figure 5a**. The BER versus the banded-matrix bandwidth (τ) at different values of the SNR

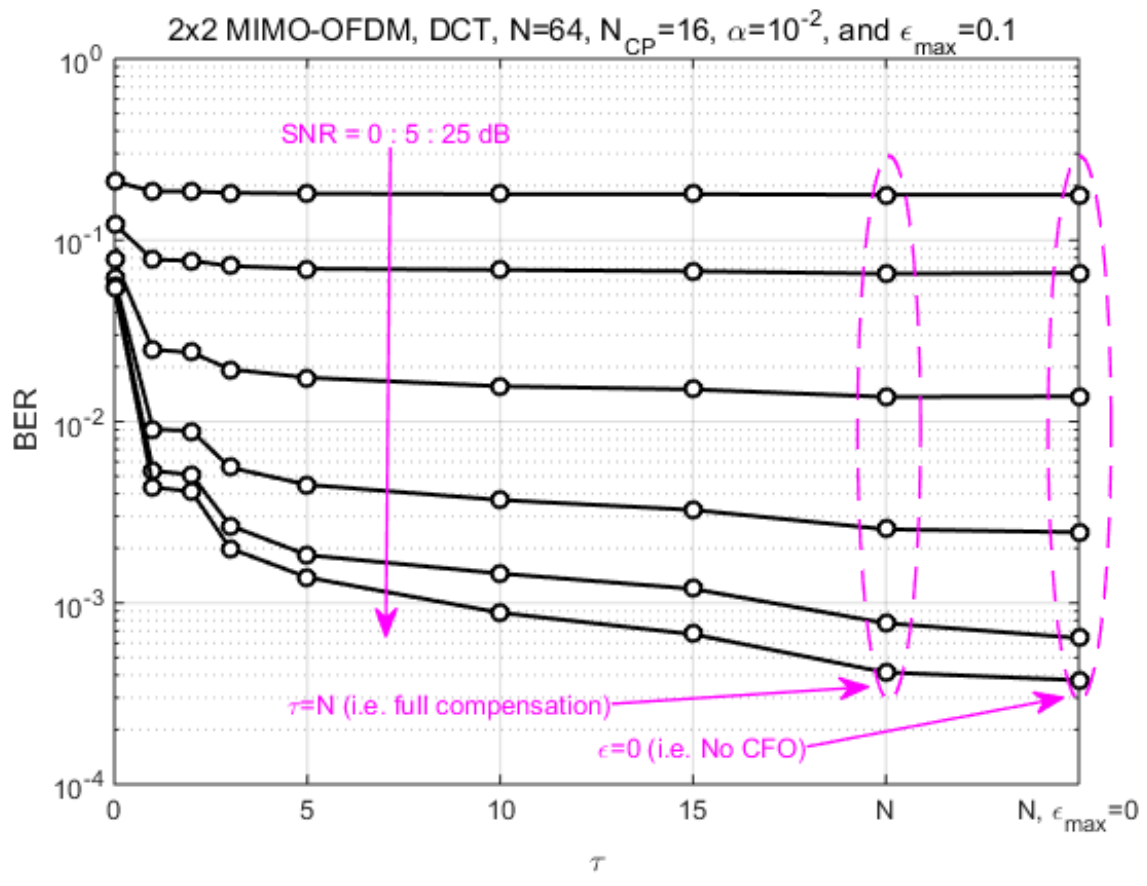

**Figure 5b**. The elevation view of Fig. 5a

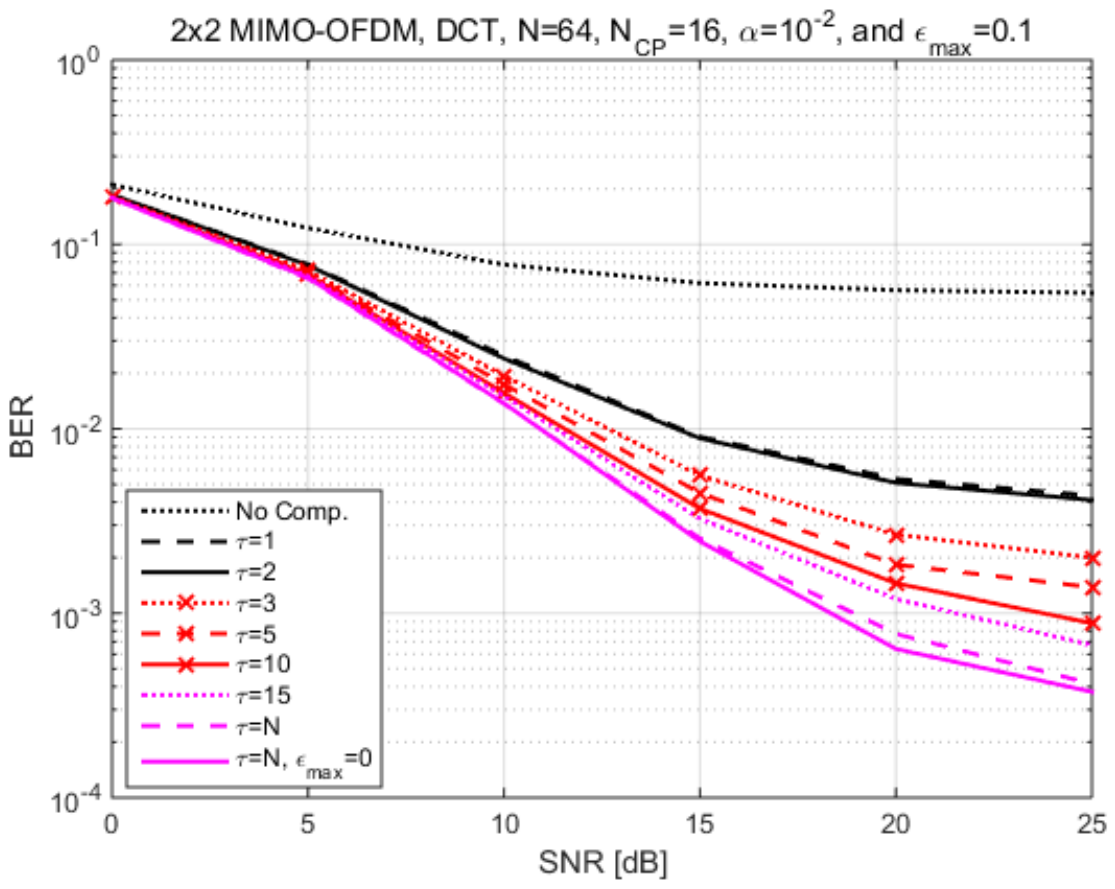

**Figure 6**. The BER versus the SNR at different values of the compensation scenarios

Figure 7a depicts the BER performance versus the normalized co-CFO at various SNR values of the proposed JLCRLZF-CDE (i.e., $\tau$=15, $\alpha$=$10^{-2}$). The elevation view of Fig. 7a is shown in Figure 7b. These figures demonstrate the significance of the compensating method in preserving BER performance at various SNR levels. As a result, let's examine the BER performance at various normalized co-CFO and compensation situations. Figure 8a depicts the BER performance with regard to the JLCRLZF-CDE at various values of the normalized co-CFO and banded-matrix bandwidth using different equalizers. Furthermore, the significant decline in BER performance happens in the absence of compensation (i.e., $\tau$=0), indicating the necessity of compensation. Figure 8b depicts the elevation view of Fig. 8a for various values of the normalized co-CFO (i.e., $\varepsilon$=0 and $\varepsilon$=0.1). It is obvious that the BER performance is saturated at $\tau \geq 5$. Figure 8c depicts the side view of Fig. 8a for various compensation situations (i.e., $\tau$=0 and $\tau$=15). It is evident that without compensation, the BER performance degrades significantly as the normalized co-CFO increases, but it improves when banded-matrix approximation compensation is used.

Figure 9 compares the BER performance of LZF and LMMSE utilizing FDE and CDE to the proposed JLCRLZF-CDE. Furthermore, FDEs necessitate an extra DFT/IDFT to accomplish the equalization procedure in the frequency domain, increasing the computational complexity. Over the variation range of the SNR, the LZF-FDE outperforms the corresponding CDE-based one. Similarly, across the variation range of the SNR, the LMMSE-FDE outperforms the corresponding CDE-based one. This appears to be a tradeoff between BER performance and computational complexity. To achieve the same BER performance as the proposed JLCRLZF-CDE at BER=$10^{-3}$, the LZF-FDE/CDE requires an additional SNR larger than 3.45 dB. On the other hand, the proposed JLCRLZF-CDE, requires an additional SNR of about 2.3 and 3.65 dB to achieve the same BER performance as LMMSE based on CDE and FDE, respectively. It should be noted that all LZF and LMMSE equalizers based on CDE/FDE are built with full matrix compensation (i.e. τ=N), whereas the proposed JLCRLZF-CDE is built with banded-matrix approximation (i.e. τ=15).

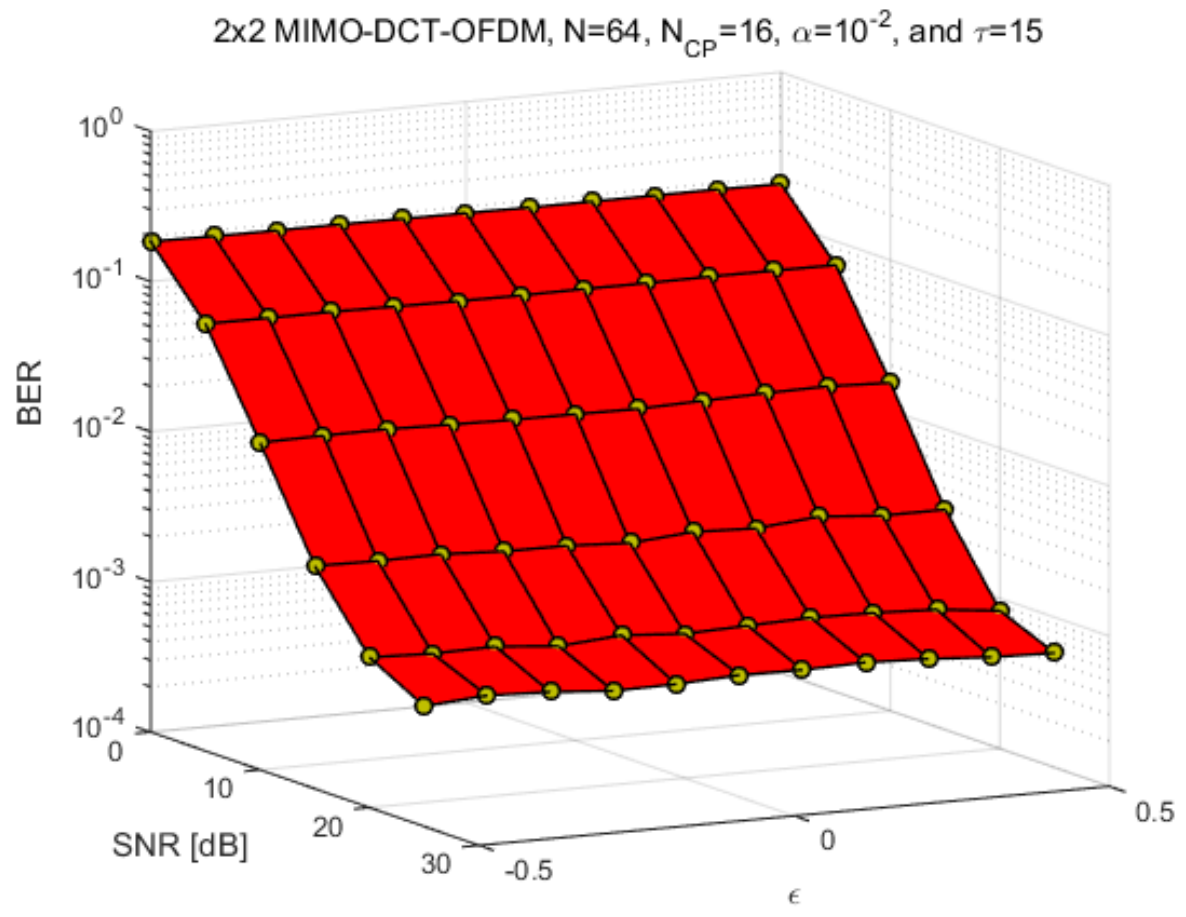

**Figure 7a**. The BER versus the normalized co-CFO at different values of the SNR

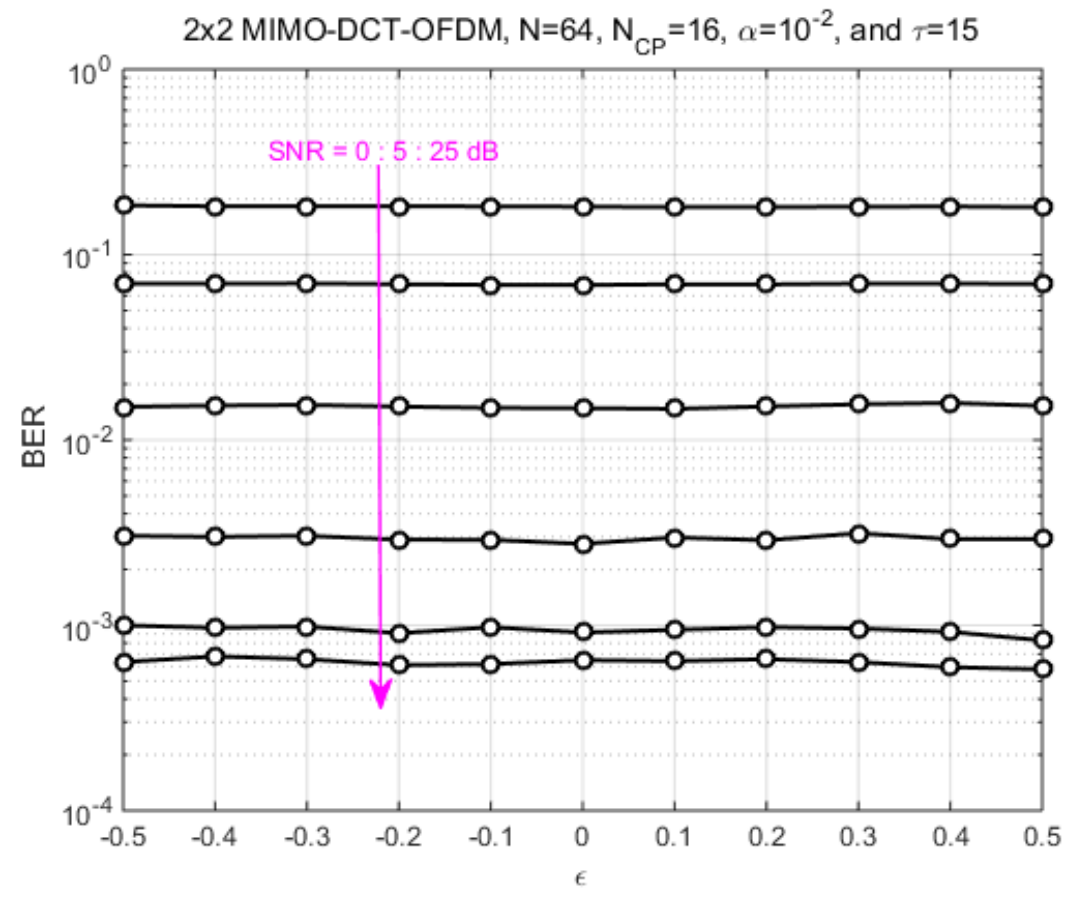

**Figure 7b**. The elevation view of Fig. 7a

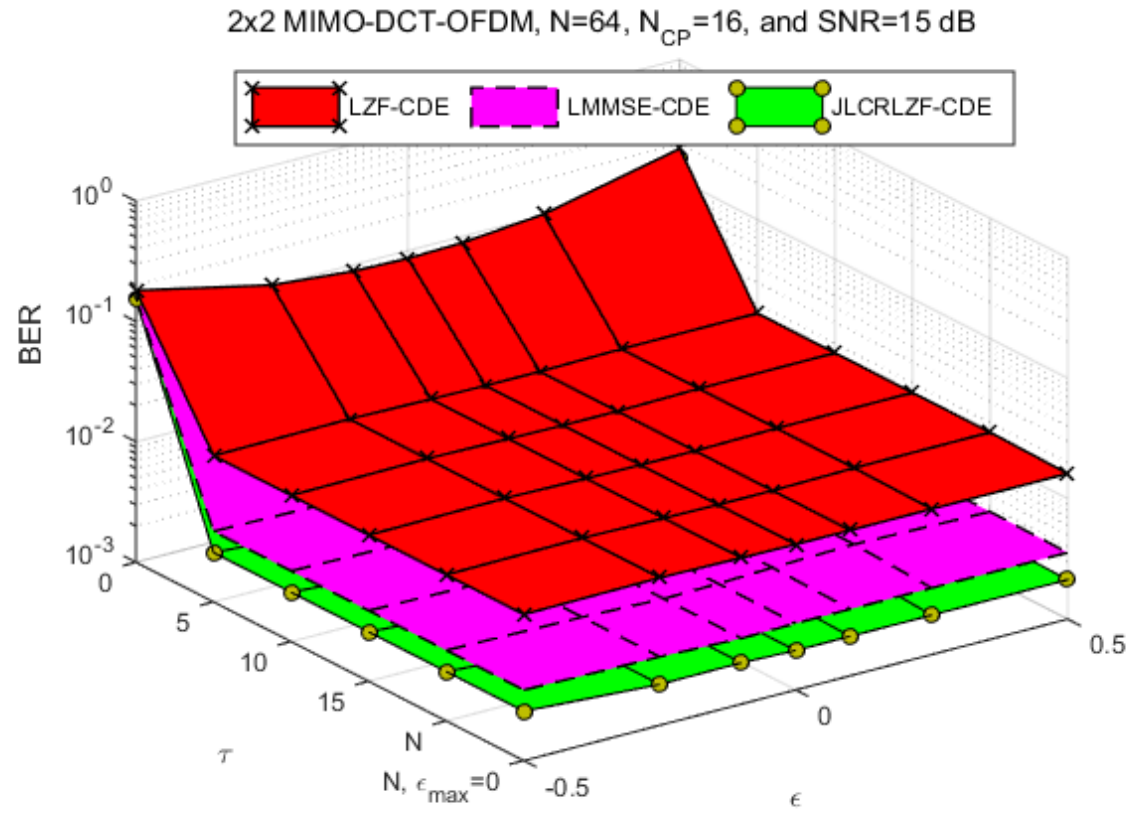

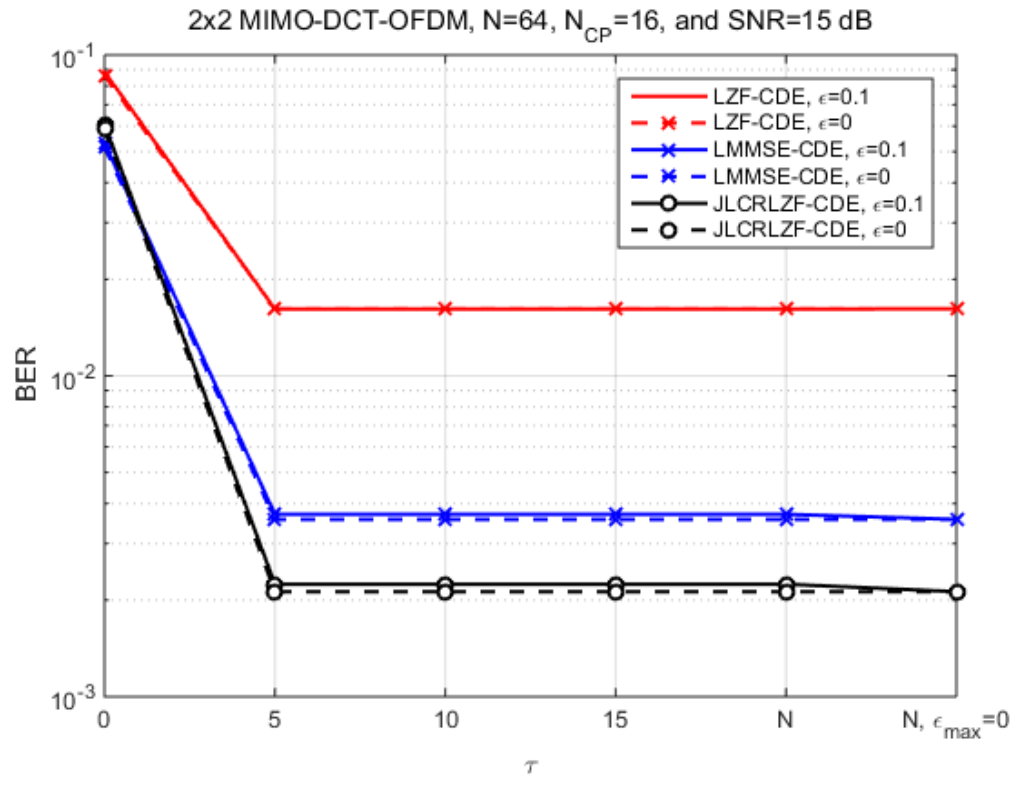

**Figure 8a**. The BER versus the normalized co-CFO at different values of the banded-matrix bandwidth

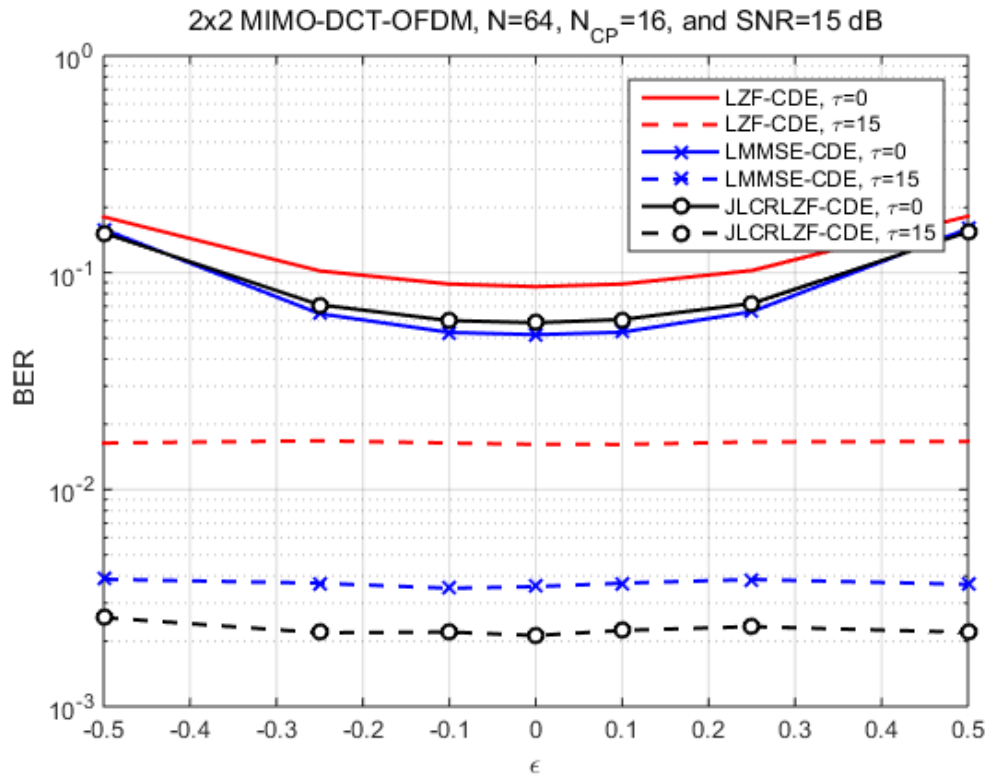

**Figure 8b**. The elevation view of Fig. 8a

**Figure 8c**. The side view of Fig. 8a

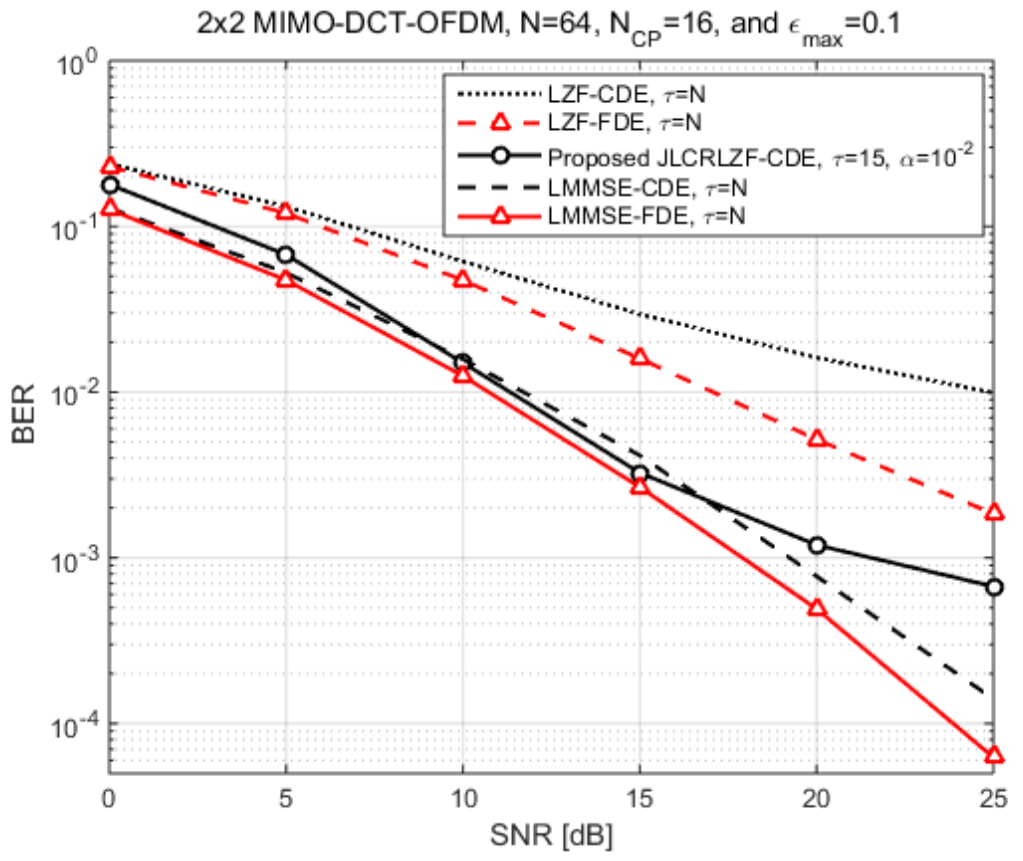

**Figure 9**. The BER versus the SNR of different linear FDEs and linear CDEs with respect to JLCRLZF-CDE.

Figure 10a depicts the BER performance versus SNR for various percentage estimation error values of the normalized co-CFO. The co-CFO is generated using a uniform random distribution in the range [-$\varepsilon_{max}$, +$\varepsilon_{max}$], and the additional percentage error of the normalized co-CFO is determined by the random value generated for the normalized co-CFO. Figure 10b depicts the BER performance versus the SNR in the presence of normalized co-CFO estimation errors of around 20%, which is the elevation view of Fig. 10a. Figure 10c depicts the BER performance versus the percentage estimation error of the normalized co-CFO, which is a side view of Fig. 10a at SNR=15 dB. These figures show the BER performance of the proposed JLCRLZF-CDE compared to other equalizers in various instances, indicating the robustness of the proposed JLCRLZF-CDE in various conditions compared to other equalizers. It should be mentioned that all LZF and LMMSE equalizers based on CDE/FDE are constructed with full matrix compensation (i.e. $\tau=N$), but the suggested JLCRLZF-CDE is produced with banded-matrix approximation (i.e. $\tau$=15).

Figure 11a illustrates the BER performance vs SNR for various estimate error levels of the channel coefficients. Figure 10b illustrates the BER performance vs SNR in the presence of channel estimation errors of $\Delta h=10^{-2}$ and $10^{-3}$, which is the elevation view of Fig. 11a. Figure 11c illustrates the BER performance vs the percentage estimation error of the channel coefficients, which is a side view of Fig. 11a at SNR=15 and 20 dB. These figures show the BER performance of the proposed JLCRLZF-CDE compared to other equalizers in various situations, showing the robustness of the proposed JLCRLZF-CDE in various conditions compared to other equalizers. All LZF and LMMSE equalizers based on CDE/FDE are built with full matrix compensation (i.e., $\tau=N$), while the proposed JLCRLZF-CDE is built with banded-matrix approximation (i.e., $\tau=15$).

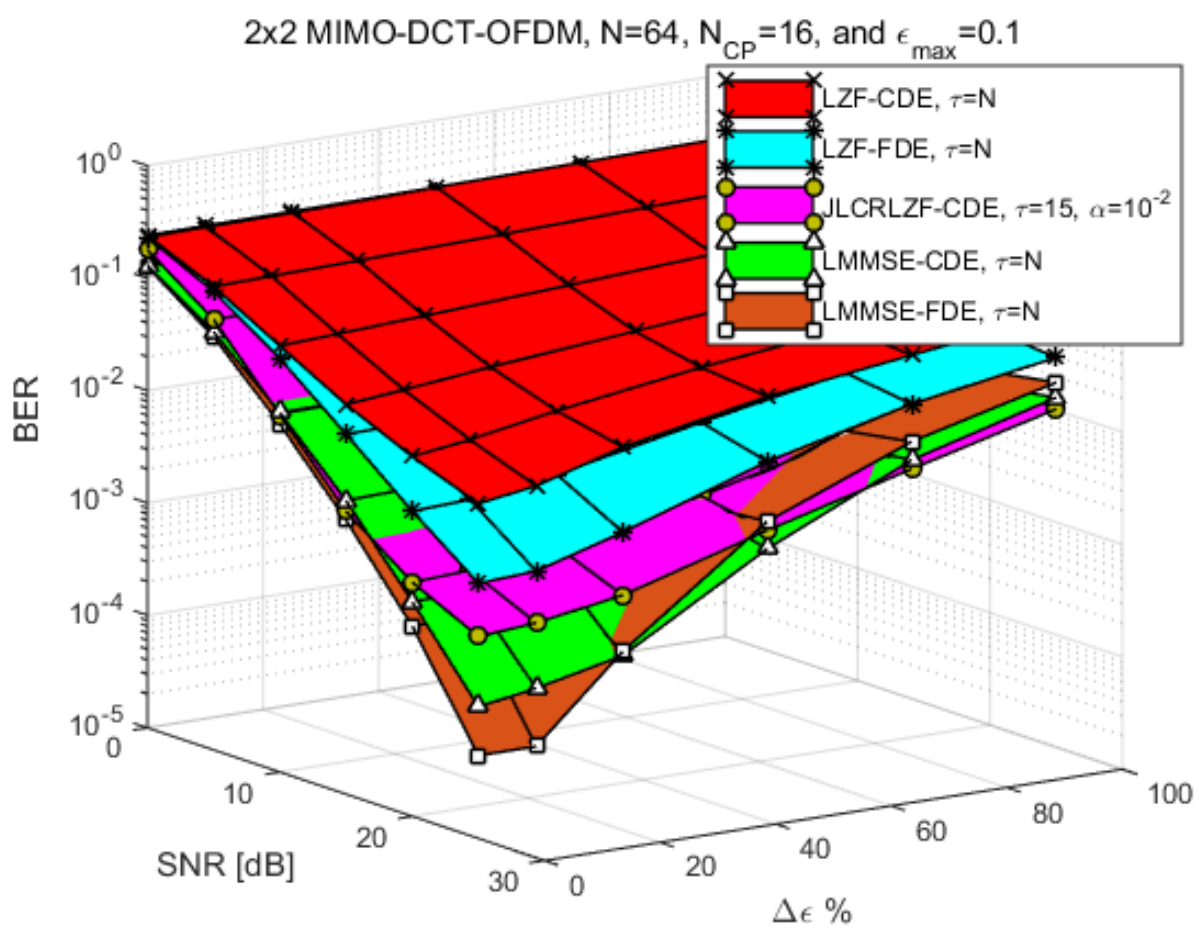

**Figure 10a**. The BER versus the estimation error percentage of the normalized co-CFO at different values of the SNR

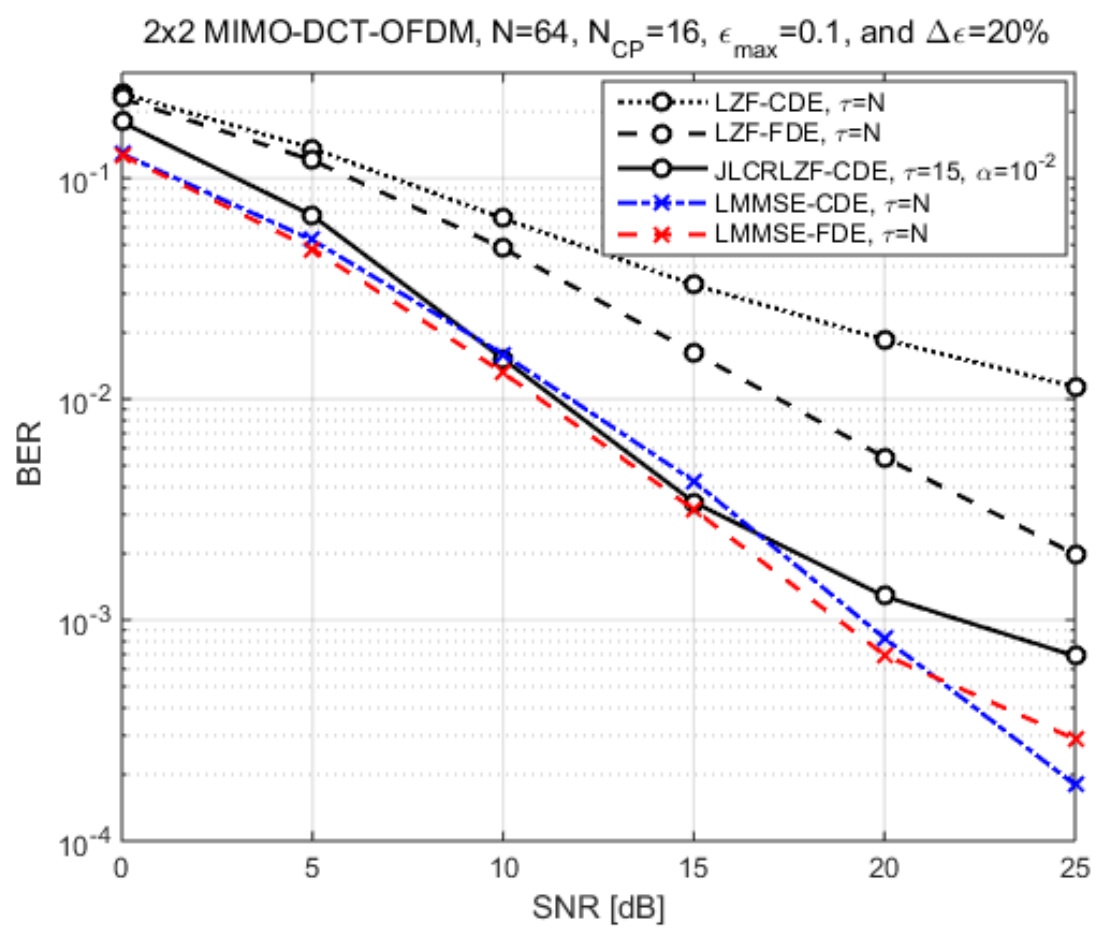

**Figure 10b**. The elevation view of Fig. 10a

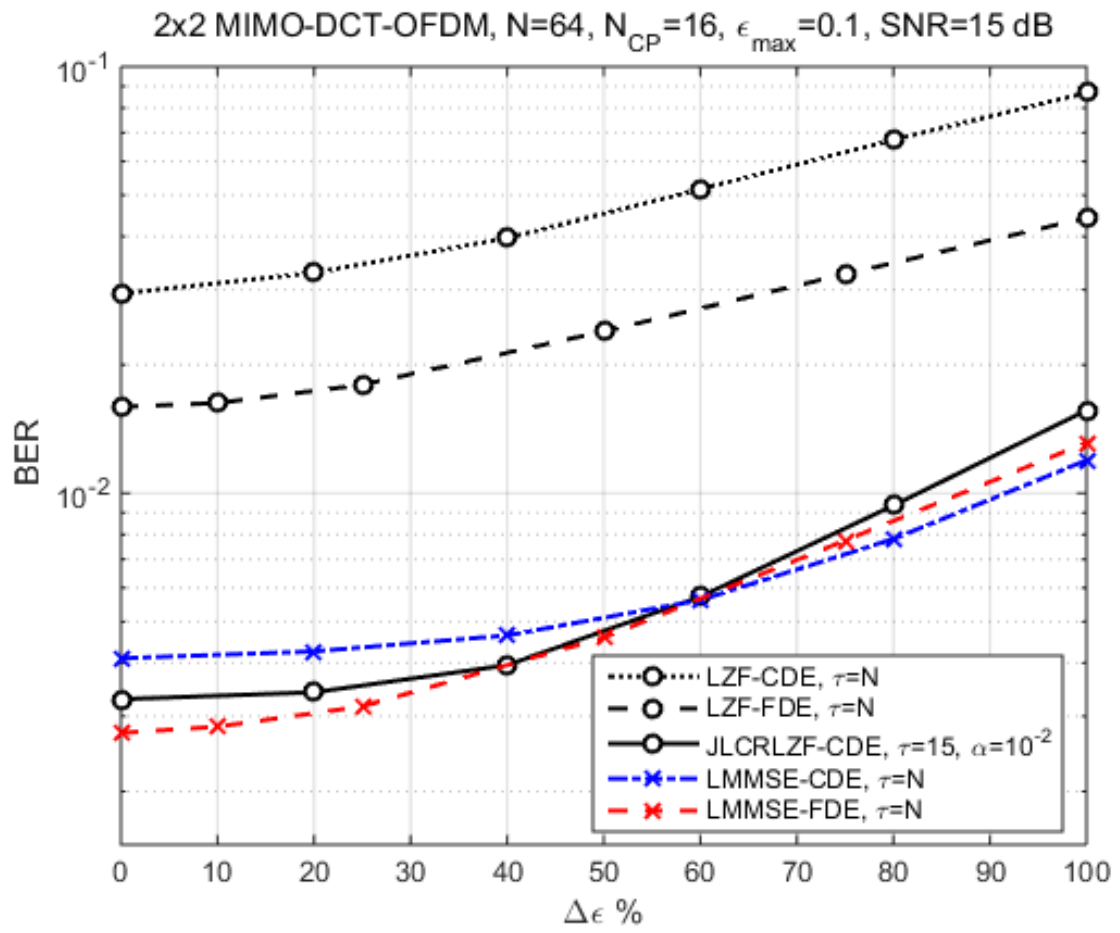

**Figure 10c**. The side view of Fig. 10a

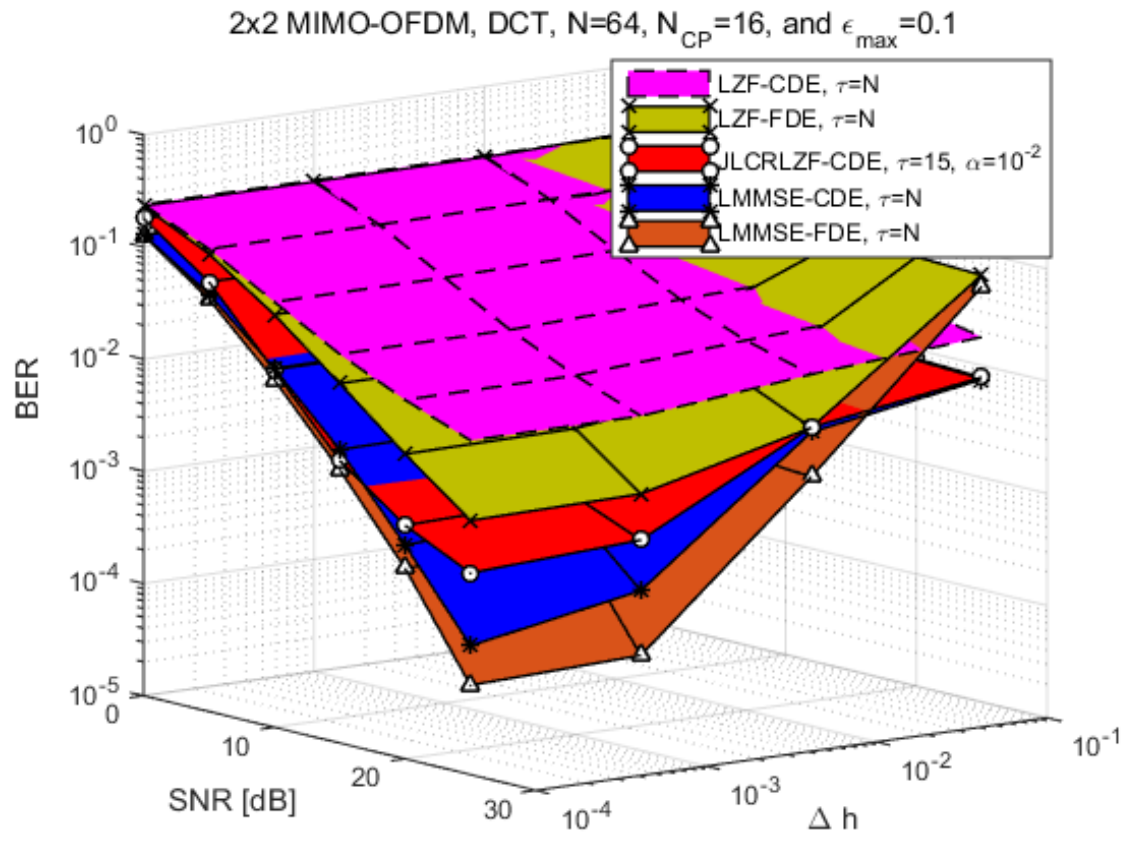

**Figure 11a**. The BER versus the channel estimation error at different values of the SNR

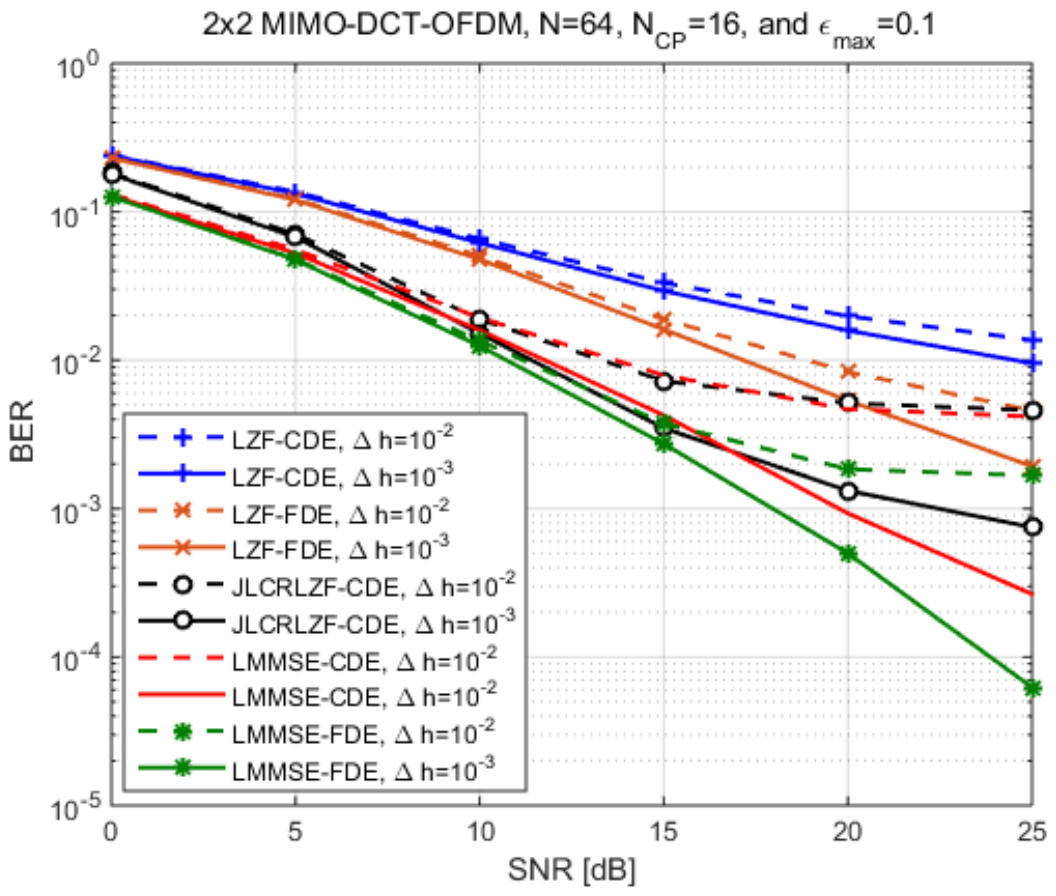

**Figure 11b**. The elevation view of Fig. 11a

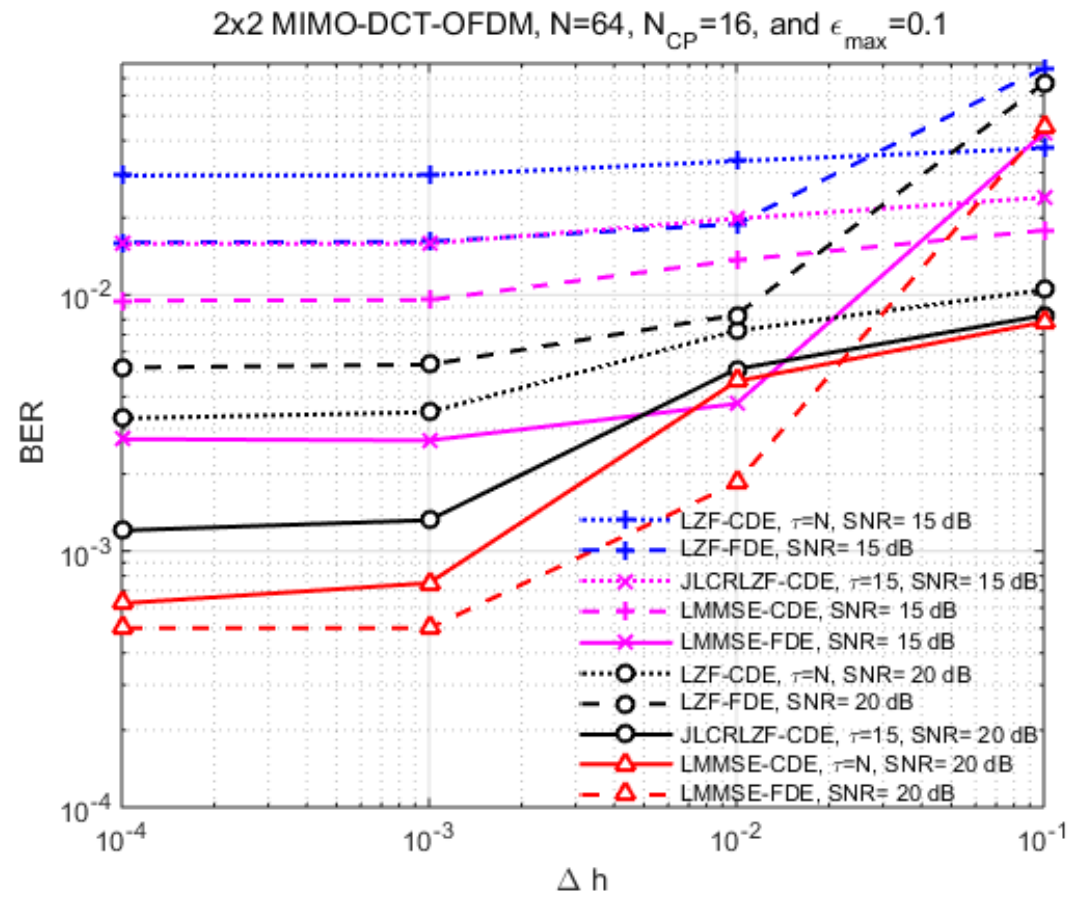

**Figure 11c**. The side view of Fig. 11a

Figure 12 shows the number of flops versus the channel configuration of the proposed JLCRLZF-CDE and other schemes for a $2^{\sigma} \times 2^{\sigma}$ MIMO-DCT-OFDM system. It is clear that the proposed JLCRLZF-CDE equalizer outperforms all schemes in terms of the number of flops. Figure 13 shows the average simulated time of various equalization procedures for different configurations of a $2^{\sigma} \times 2^{\sigma}$ MIMO-DCT-OFDM system, which shows that the proposed JLCRLZF-CDE equalizer outperforms these schemes and needs lower simulated time, which can be expressed as a reduction in simulation time. Figure 14 shows the simulated reduction time of various equalization procedures for different configurations of a $2^{\sigma} \times 2^{\sigma}$ MIMO-DCT-OFDM system compared to that of the proposed JLCRLZF-CDE. Note that, $\sigma = \{1,2,3,4, \ldots\}$.

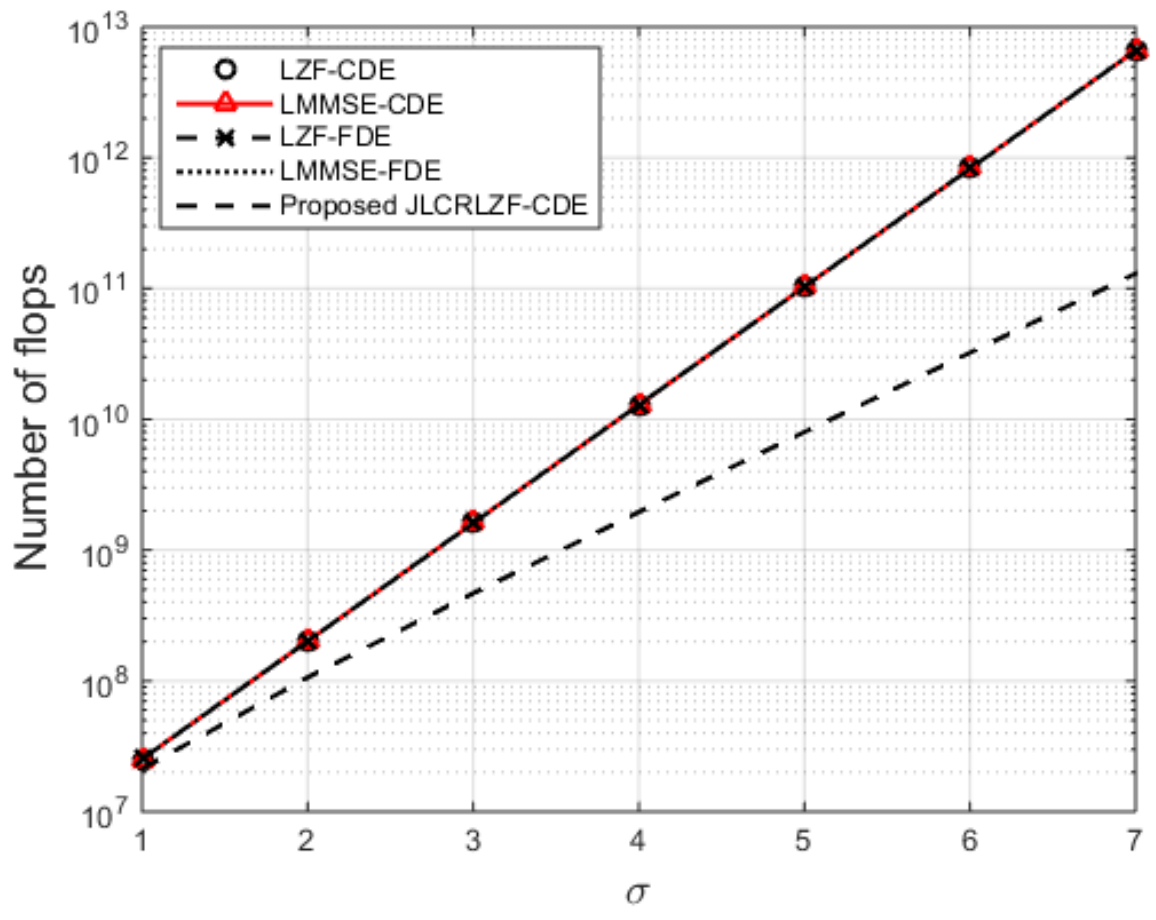

**Figure 12**. The number of flops versus the channel configurations

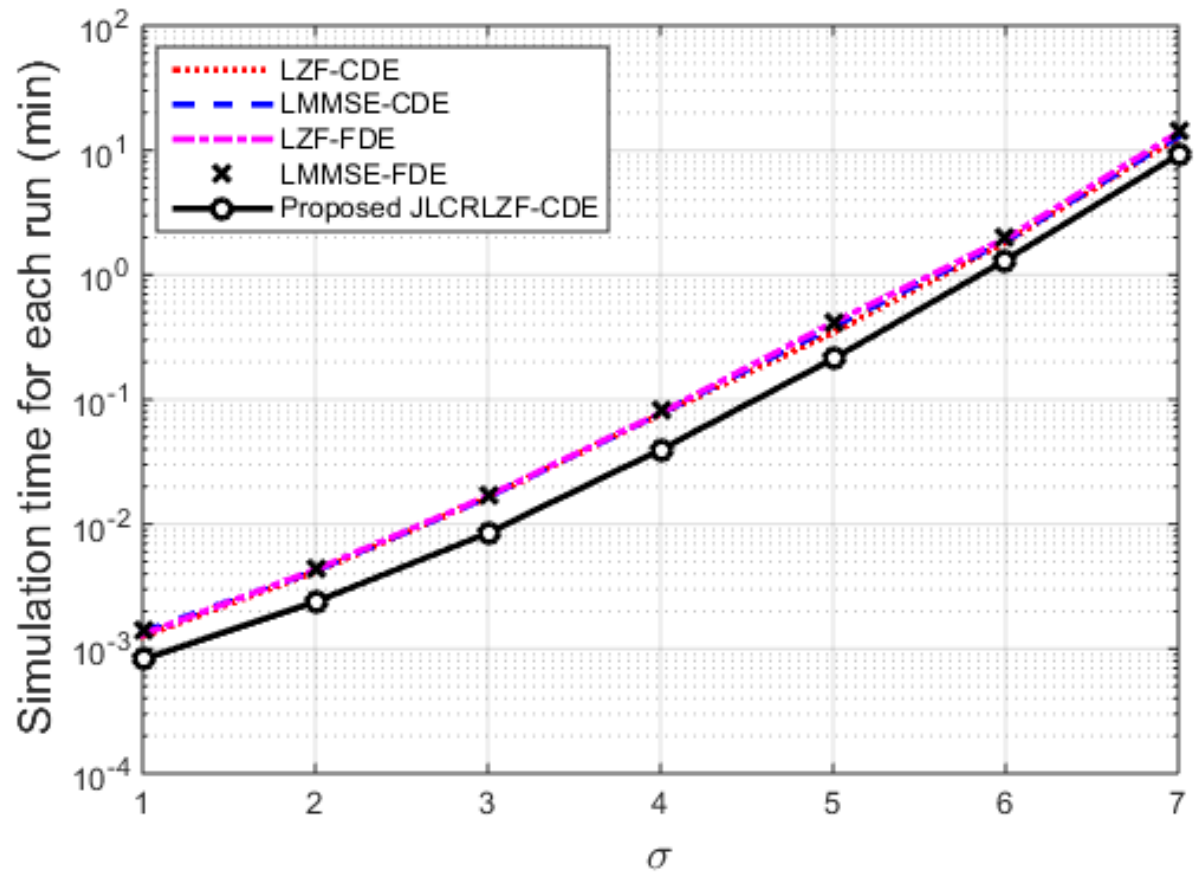

**Figure 13**. The simulation time versus the channel configuration**s**

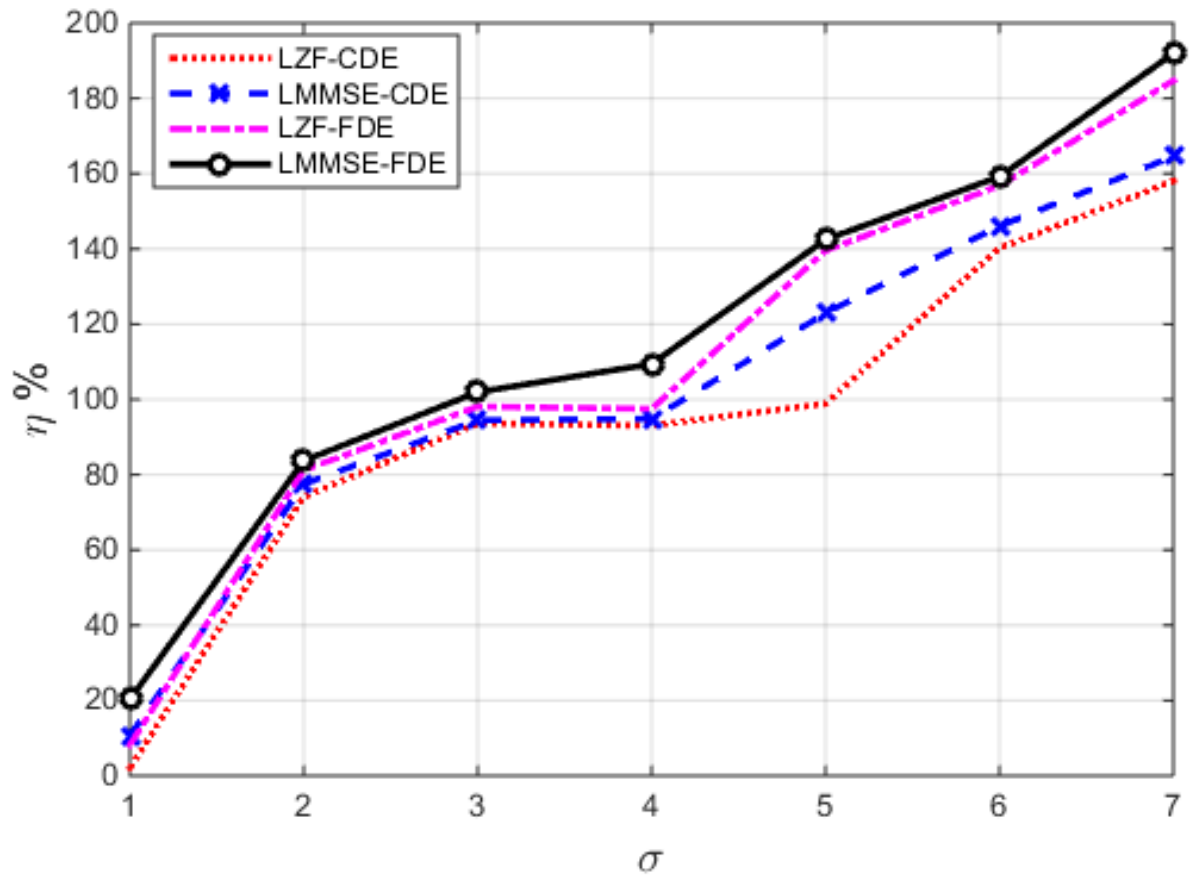

**Figure 14**. The simulation time reduction versus the channel configuration**s**

## V. Complexity Analysis

The complexity evaluation of several equalization and compensating algorithms is provided in this section. Real multiplication/addition/division is counted as one operation that may be performed using half flops [21]. The term "flops" refers to the number of floating-point operations performed per second. Table 2 shows the total number of operations and flops associated with various mathematical operation scenarios. Table 3 shows the total number of operation and flops associated with various full-matrix operation scenarios. Table 4 shows the total number of operation and flops associated with various banded-matrix operation scenarios. So, for each equalization described later, let's calculate the total number of operations and flops for different configuration orders like 2×2, 4×4, 8×8, 16×16, 32×32, 64×64, and 128×128 MIMO-DCT-OFDM system.

*A. The LZF-CDE*

The solution matrix of the conventional LZF-CDE (τ=*N*) can be written as follows:

$$\mathbf{\Upsilon}_{\text{LZF-CDE}} = (\mathbf{\Pi}^{H}\mathbf{\Pi})^{-1}\,\mathbf{\Pi}^{H} \tag{21}$$

where $\mathbf{\Pi} \in \mathbb{C}^{M\times M}, M = 2^{\sigma}N$, as $2^{\sigma} \times 2^{\sigma}$ MIMO-DCT-OFDM system, $\sigma$ =1, 2, 3, 4, 5, 6, and 7. according to Eq. (21), the LZF-CDE involves the creation of two complex matrices through matrix multiplication and matrix inversion. According to Tables 3, and 4 multiplying two complex matrices needs $16M^3$ operations. $8M^3$ flops. $8M^3 + 2M^2 + M$ operations are required for complex matrix inversion, which equates to $4M^3 + M^2 + M/2$ flops. The Singular Value Decomposition (SVD) idea [22] can be used to validate the matrix inversion. The LZF-CDE solution matrix is multiplied by the received vector, which requires $8M^2$ operations and equates to $4M^2$ flops. As a result, the LZF-CDE requires $12M^3 + 5M^2 + M/2$ flops. The noise enhancement problem caused by direct matrix inversion is the major general difficulty of the LZF equalization. According to the simulation parameters provided in Table 1, and utilizing a laptop with a core i5 CPU, the LZF-CDE takes 0.337, 4.142, 16.442, 76.323, 344.851, 1831, 12570.318 m min for $\sigma$ =1, 2, 3, 4, 5, 6, 7, respectively to complete its procedure on each run.

*B. The LMMSE-CDE*

The solution matrix of the conventional LMMSE-CDE (τ=*N*) can be written as follows:

$$\mathbf{\Upsilon}_{\text{LMMSE-CDE}} = \left(\mathbf{\Pi}^{H}\mathbf{\Pi} + \frac{\mathrm{R}_{\mathbf{Z}}}{\sigma_{\mathbf{X}}^{2}}\,\mathbf{I}_{2N\times 2N}\right)^{-1}\,\mathbf{\Pi}^{H} \tag{22}$$

where $\sigma_X^2$ is the transmitted signal power, $\mathrm{R}_{\mathbf{X}} = \mathbb{E}\{\mathbf{Z}\,.\,\mathbf{Z}^{H}\}$ is the AWGN covariance matrix, $\mathbb{E}\{\#\}$ is the expectation of #, and the term $\left(\frac{\sigma_{\mathbf{X}}^{2}}{\mathrm{R}_{\mathbf{Z}}}\right)$ is the SNR value. The difference between Eqs. (21, 22) is the estimate of the SNR vales, which may have resulted in a time delay processing as compared to LZF-CDE. As a result, the LMMSE-CDE requires $12M^3 + 5M^2 + 2.5M$ flops in addition to the calculation of the SNR value. According to the simulation parameters provided in Table 1, and utilizing a laptop with a core i5 CPU, the LMMSE-CDE takes 0.365, 4.223, 16.512, 74.806, 386.557, 1876.315, and 12888.468 m min $\sigma$ =1, 2, 3, 4, 5, 6, 7, respectively to complete its procedure on each run.

*C. The LZF-FDE*

In fact, the frequency domain equalization method necessitates an additional DFT/IDFT for each branch. In the case of a $2^{\sigma} \times 2^{\sigma}$ MIMO-DCT-OFDM system, we employ two DFT/IDFT blocks. According to [23], when the data vector length is $N \times 1$, the DFT or IDFT requires $N^2$ multiplications and $N(N-1)$ additions. This is equivalent to $2N^2 - N$ operations, which are equivalent to $N^2 - N/2$ flops. As a result, two DFT/IDFT blocks necessitate $M^2 - M$ flops. Thus, the LZF-FDE requires $12M^3 + 6M^2 - M/2$ flops. According to the simulation parameters provided in Table 1, and utilizing a laptop with a core i5 CPU, the LZF-FDE takes 0.357, 4.311, 16.820, 78.084, 415.417, 1958.331, and 13870.996 m min, respectively to complete its procedure on each run.

*D. The LMMSE-FDE*

In the same manner, the LMMSE-FDE requires $12M^3 + 6M^2 + 2M$ flops in addition to the calculation of the SNR value. According to the simulation parameters provided in Table 1, and utilizing a laptop with a core i5 CPU, the LMMSE-FDE takes 0.397, 4.371, 17.146, 82.794, 420.496, 1976.542, and 14220.691 m min, respectively to complete its procedure on each run.

*E. The Proposed JLCRLZF-CDE*

According to Eq. (19), the JLCRLZF-CDE involves the creation of two complex matrices through matrix multiplication and matrix inversion.

The first complex matrix via matrix multiplication $(\text{i.e. } \boldsymbol{\mu}^H\boldsymbol{\mu})$, where the $\boldsymbol{\mu}$ matrix consists of $2^{2\sigma}$ banded sub-matrices, each with a bandwidth of τ, will be multiplied by the $\boldsymbol{\mu}^H$, which likewise has a bandwidth of τ. As a result, the output will be $2^{2\sigma}$ banded sub-matrices, each with a bandwidth of 2τ [16].

According to [16], and Table 4 each output banded-matrix requires around $N(32\tau^2+16\tau+8)$ operations, which equates to $N(16\tau^2+8\tau+4)$ flops. As a result, the creation of a $\boldsymbol{\mu}^H\boldsymbol{\mu}$ matrix necessitates the use of about $2^{2\sigma}N(16\tau^2+8\tau+4)$ flops with a bandwidth of 2τ for each sub-matrix. The output matrix is combined with the regularization parameter, which requires $2^{\sigma}N$ flops. This is followed by complex matrix inversion of 2τ bandwidth. In what follows, a detailed count of the arithmetic in the case of 2×2 MIMO-DCT-OFDM system, $\sigma=1$ is introduced, then higher order configuration is developed.

*i.* 2×2 *MIMO-DCT-OFDM system*

The matrix inversion in the case of 2×2 MIMO-DCT-OFDM system consists of four sub-matrices $(\text{i.e.,}\ 2^1\times 2^1=4)$, which can be written as [24]:

$$(\boldsymbol{\mu}^H\boldsymbol{\mu}+\alpha\,\mathbf{I}_{2N\times 2N})^{-1}=\begin{bmatrix}\overbrace{\underbrace{\begin{matrix}\lambda_{1,1}\\ \lambda_{2,1}\end{matrix}}_{\lambda_3}}^{\lambda_1} & \overbrace{\underbrace{\begin{matrix}\lambda_{1,2}\\ \lambda_{2,2}\end{matrix}}_{\lambda_4}}^{\lambda_2}\end{bmatrix}^{-1}=\begin{bmatrix}\bar{\lambda}_1 & \bar{\lambda}_2\\ \bar{\lambda}_3 & \bar{\lambda}_4\end{bmatrix}=\begin{bmatrix}\Phi & -\Phi\lambda_2{\lambda_4}^{-1}\\ -{\lambda_4}^{-1}\lambda_3\Phi & {\lambda_4}^{-1}+{\lambda_4}^{-1}\lambda_3\Phi\lambda_2{\lambda_4}^{-1}\end{bmatrix} \quad (23)$$

where, $\Phi=\left(\lambda_1-\lambda_2{\lambda_4}^{-1}\lambda_3\right)^{-1}, \lambda_i, i\in\{1,2,3,4\}$ has $N\times N$ dimensions and a bandwidth of 2τ. Let's figure out how many flops are required for the $\bar{\lambda}_i, i\in\{1,\ 2,3,\ 4\}$ structure. Table 4 shows that the ${\lambda_4}^{-1}$ matrix requires $N(40\tau^2+42\tau+1)$ operations, which equates to $N(20\tau^2+21\tau+0.5)$ flops and a bandwidth of 2τ. The $\lambda_2{\lambda_4}^{-1}$ matrix requires about $N(256\tau^2+32\tau+4)$ flops with a bandwidth of 4τ, which is limited to 2τ. In the similar way, $\lambda_2{\lambda_4}^{-1}\lambda_3$ building necessitates roughly $N(256\tau^2+32\tau+4)$ flops of 4τ bandwidth, which is limited to 2τ, whereas matrix subtraction necessitates $N(4\tau+0.5)$ flops. Now, matrix inversion is used to create Φ with a bandwidth of 2τ. This is equivalent to $N(20\tau^2+21\tau+0.5)$ flops. As a result, the Φ matrix construction requires about $N(552\tau^2+110\tau+9.5)$ flops. In the same manner, the computational complexity is decreased by limiting the bandwidth of each sub-matrix. Thus, we can constrain the $\bar{\lambda}_i$ matrices to 2τ bandwidth, which has no discernible effect on BER performance. The $\bar{\lambda}_2$ construction needs a complex matrix by matrix multiplication of the $\Phi$ matrix, and the constructed matrix later $\left(\text{i.e., } \lambda_2\lambda_4^{-1}\right)$. This equates to $N(64\tau^2 + 32\tau + 4)$ flops of $4\tau$ bandwidth, which is limited to $2\tau$. The $\bar{\lambda}_3$ construction needs complex matrix by matrix multiplication of $\lambda_4^{-1}\lambda_3$ and $\Phi$. Firstly, the construction of the $\lambda_4^{-1}\lambda_3$ matrix needs $N(64\tau^2 + 32\tau + 4)$ flops of $4\tau$ bandwidth, which is limited to $2\tau$. This is followed by the multiplication of the $\Phi$ matrix, which needs $N(64\tau^2 + 32\tau + 4)$ flops of $4\tau$ bandwidth, which is limited to $2\tau$. Thus, the construction of the $\bar{\lambda}_3$ matrix needs $N(128\tau^2 + 64\tau + 8)$ flops of $4\tau$ bandwidth, which is limited to $2\tau$. In fact, $\bar{\lambda}_4 = \lambda_4^{-1} + \bar{\lambda}_3\,\lambda_2\,\lambda_4^{-1}$, which equates to $N(64\tau^2 + 36\tau + 4.5)$ flops. The creation of $\bar{\lambda}_i$ matrices require roughly $N(680\tau^2 + 274\tau + 44)$ flops in total.

The generated matrix$(\text{i.e., } (\boldsymbol{\mu}^H\boldsymbol{\mu} + \alpha\,\mathbf{I}_{2N\times 2N})^{-1})$ of $2\tau$ limited bandwidth is multiplied by the $\boldsymbol{\mu}^H$ matrix to get the JLCRLZF-CDE solution matrix stated in Eq. (19). This is equivalent to $2^{2\sigma}N(144\tau^2 + 24\tau + 4)$ flops, $\sigma = 1$ with $3\tau$ bandwidth, which is limited also to $2\tau$ bandwidth. As a result, the JLCRLZF-CDE solution matrix requires roughly $N(1448\tau^2 + 370\tau + 60)$ flops to create. The receiving vector is multiplied by the JLCRLZF-CDE solution matrix, which requires $2^{2\sigma}N(16\tau + 8)$ flops. As a result, the JLCRLZF-CDE necessitates $N(1448\tau^2 + 434\tau + 92)$ flops. The JLCRLZF-CDE takes 0.330 m min to complete its operation on each run, according to the simulation parameters in Table 1 and using a laptop with a core i5 CPU.

*i.* 4×4 *MIMO-DCT-OFDM system*

The matrix inversion in the case of 4×4 MIMO-DCT-OFDM system consists of 16 sub-matrices $(\text{i.e., } 2^2 \times 2^2 = 16)$, which can be written as:

$$(\boldsymbol{\mu}^H\boldsymbol{\mu} + \alpha\, \mathbf{I}_{4N\times 4N})^{-1} = \begin{bmatrix} \overbrace{\boxed{\begin{matrix} \lambda_{1,1} & \lambda_{1,2} \\ \lambda_{2,1} & \lambda_{2,2} \end{matrix}}}^{\lambda_1} & \overbrace{\boxed{\begin{matrix} \lambda_{1,3} & \lambda_{1,4} \\ \lambda_{2,3} & \lambda_{2,4} \end{matrix}}}^{\lambda_2} \\ \underbrace{\boxed{\begin{matrix} \lambda_{3,1} & \lambda_{3,2} \\ \lambda_{4,1} & \lambda_{4,2} \end{matrix}}}_{\lambda_3} & \underbrace{\boxed{\begin{matrix} \lambda_{3,3} & \lambda_{3,4} \\ \lambda_{4,3} & \lambda_{4,4} \end{matrix}}}_{\lambda_4} \end{bmatrix}^{-1} = \begin{bmatrix} \bar{\lambda}_1 & \bar{\lambda}_2 \\ \bar{\lambda}_3 & \bar{\lambda}_4 \end{bmatrix} \tag{24}$$

$$= \begin{bmatrix} \Phi & -\Phi\lambda_2{\lambda_4}^{-1} \\ -{\lambda_4}^{-1}\lambda_3\Phi & {\lambda_4}^{-1} + {\lambda_4}^{-1}\lambda_3\Phi\lambda_2{\lambda_4}^{-1} \end{bmatrix}$$

*Construction of $\bar{\lambda}_1$ matrix*

$\bar{\lambda}_1 = \Phi = \left(\lambda_1 - \lambda_2{\lambda_4}^{-1}\lambda_3\right)^{-1}, \lambda_i, i \in \{1,2,3,4\}$ has $4N \times 4N$ dimensions, which consists of four sub-matrices each with a bandwidth of 2τ. Let's figure out how many flops are required for the $\bar{\lambda}_i, i \in \{1,\ 2,3,\ 4\}$ structure. In fact, the ${\lambda_4}^{-1}$ consists of four sub-matrices each with bandwidth of 2τ, which can construct using the same steps of the 2×2 system configuration described in Eq. (23). It is should note that, each constructed matrix exceeds 2τ bandwidth is limited to 2τ bandwidth. Thus, the ${\lambda_4}^{-1}$ matrix construction needs $N(808\tau^2 + 242\tau + 26)$ flops, which consists of four sub-matrices each has a 2τ bandwidth.

The $\lambda_2{\lambda_4}^{-1}$ consists of four sub-matrices of limited 2τ bandwidth with $N(1024\tau^2 + 128\tau + 16)$ flops. In the similar way, $\lambda_2{\lambda_4}^{-1}\lambda_3$ building necessitates roughly $N(1024\tau^2 + 128\tau + 16)$ flops of limited 2τ bandwidth for each of the four sub-matrices. The matrix subtraction necessitates $N(16\tau + 2)$ flops.

Now, matrix inversion is used to create Φ with a bandwidth of 2τ. This is equivalent to $N(808\tau^2 + 242\tau + 26)$ flops. As a result, the Φ matrix construction requires about $N(3664\tau^2 + 756\tau + 86)$ flops.

*Construction of $\bar{\lambda}_2$ matrix*

The $\bar{\lambda}_2$ construction needs a complex matrix by matrix multiplication of the Φ matrix, and the constructed matrix later $\left(\text{i.e., } \lambda_2{\lambda_4}^{-1}\right)$, moreover each of them $\left(\text{i.e., } \lambda_2{\lambda_4}^{-1},\ \Phi\right)$ consists of four sub-matrices. This equates to $N(256\tau^2 + 128\tau + 16)$ flops of limited 2τ bandwidth.

*Construction of $\bar{\lambda}_3$ matrix*

The $\bar{\lambda}_3$ construction needs complex matrix by matrix multiplication of ${\lambda_4}^{-1}\lambda_3$ and $\Phi$. Firstly, the construction of the ${\lambda_4}^{-1}\lambda_3$ matrix needs $N(256\tau^2 + 128\tau + 16)$ flops of limited 2τ bandwidth. This is followed by the multiplication of the Φ matrix, which needs $N(256\tau^2 + 128\tau + 16)$ flops of limited 2τ bandwidth. Thus, the construction of the $\bar{\lambda}_3$ matrix needs $N(512\tau^2 + 256\tau + 32)$ flops of limited to 2τ bandwidth.

*Construction of $\bar{\lambda}_4$ matrix*

In fact, $\bar{\lambda}_4 = {\lambda_4}^{-1} + \bar{\lambda}_3\, \lambda_2\, {\lambda_4}^{-1}$ , which equates to $N(256\tau^2 + 144\tau + 18)$ flops. The creation of $\bar{\lambda}_i$ matrices require roughly $N(4688\tau^2 + 1284\tau + 152)$ flops in total. Note that, the construction of the $(\boldsymbol{\mu}^H\boldsymbol{\mu} + \alpha\,\mathbf{I}_{4N\times 4N})$ matrix requires $2^{2\sigma}N(16\tau^2 + 8\tau + 4)$+$2^{\sigma}N$ flops, $\sigma = 2$.

The generated matrix(i.e., $(\boldsymbol{\mu}^H\boldsymbol{\mu} + \alpha\,\mathbf{I}_{4N\times 4N})^{-1}$) of 2τ limited bandwidth is multiplied by the $\boldsymbol{\mu}^H$ matrix to get the JLCRLZF-CDE solution matrix stated in Eq. (19). This is equivalent to $2^{2\sigma}N(144\tau^2 + 24\tau + 4)$ flops, with 3τ bandwidth, which is limited also to 2τ bandwidth. As a result, the JLCRLZF-CDE solution matrix requires roughly $N(7248\tau^2 + 1796\tau + 284)$ flops to create. The receiving vector is multiplied by the JLCRLZF-CDE solution matrix, which requires $2^{2\sigma}N(16\tau + 8)$ flops. As a result, the JLCRLZF-CDE necessitates $N(7248\tau^2 + 2052\tau + 412)$ flops. The JLCRLZF-CDE takes 2.379 m min to complete its operation on each run, according to the simulation parameters in Table 1 and using a laptop with a core i5 CPU.

*ii.* 8×8 *MIMO-DCT-OFDM system*

The matrix inversion in the case of 8×8 MIMO-DCT-OFDM system consists of 64 sub-matrices (i.e., $2^3 \times 2^3 = 64$), which can be written as:

$$(\boldsymbol{\mu}^H\boldsymbol{\mu} + \alpha\,\mathbf{I}_{8N\times 8N})^{-1} = \begin{bmatrix} \overbrace{\boxed{\begin{matrix} \lambda_{1,1} & \lambda_{1,2} & \cdots & \lambda_{1,4} \\ \lambda_{2,1} & \lambda_{2,2} & \cdots & \lambda_{2,4} \\ \vdots & \vdots & \ddots & \vdots \\ \lambda_{4,1} & \lambda_{4,2} & \cdots & \lambda_{4,4} \end{matrix}}}^{\lambda_1} & \overbrace{\boxed{\begin{matrix} \lambda_{1,5} & \lambda_{1,6} & \cdots & \lambda_{1,8} \\ \lambda_{2,5} & \lambda_{2,6} & \cdots & \lambda_{2,8} \\ \vdots & \vdots & \ddots & \vdots \\ \lambda_{4,5} & \lambda_{4,6} & \cdots & \lambda_{4,8} \end{matrix}}}^{\lambda_2} \\ \underbrace{\boxed{\begin{matrix} \lambda_{5,1} & \lambda_{5,2} & \cdots & \lambda_{5,4} \\ \lambda_{6,1} & \lambda_{6,2} & \cdots & \lambda_{6,4} \\ \vdots & \vdots & \ddots & \vdots \\ \lambda_{8,1} & \lambda_{8,2} & \cdots & \lambda_{8,4} \end{matrix}}}_{\lambda_3} & \underbrace{\boxed{\begin{matrix} \lambda_{5,5} & \lambda_{5,6} & \cdots & \lambda_{5,8} \\ \lambda_{6,5} & \lambda_{6,6} & \cdots & \lambda_{6,8} \\ \vdots & \vdots & \ddots & \vdots \\ \lambda_{8,5} & \lambda_{8,6} & \cdots & \lambda_{8,8} \end{matrix}}}_{\lambda_4} \end{bmatrix}^{-1} = \begin{bmatrix} \bar{\lambda}_1 & \bar{\lambda}_2 \\ \bar{\lambda}_3 & \bar{\lambda}_4 \end{bmatrix} \tag{25}$$

$$= \begin{bmatrix} \Phi & -\Phi\lambda_2{\lambda_4}^{-1} \\ -{\lambda_4}^{-1}\lambda_3\Phi & {\lambda_4}^{-1} + {\lambda_4}^{-1}\lambda_3\Phi\lambda_2{\lambda_4}^{-1} \end{bmatrix}$$

*Construction of $\bar{\lambda}_1$ matrix*

$\bar{\lambda}_1 = \Phi = \left(\lambda_1 - \lambda_2{\lambda_4}^{-1}\lambda_3\right)^{-1}, \lambda_i, i \in \{1, 2, 3, 4\}$ has $8N \times 8N$ dimensions, which consists of sixteen sub-matrices each with a bandwidth of 2τ.

Let's figure out how many flops are required for the $\bar{\lambda}_i, i \in \{1,\ 2, 3,\ 4\}$ structure. In fact, the ${\lambda_4}^{-1}$ consists of sixteen sub-matrices each with bandwidth of 2τ, which can construct using the same steps of the 2×2, and 4×4 system configuration described in Eqs. (23), and (24), respectively. It is should note that, each constructed matrix exceeds 2τ bandwidth is limited to 2τ bandwidth. Thus, the ${\lambda_4}^{-1}$ matrix construction needs $N(4688\tau^2 + 1284\tau + 152)$ flops, which consists of sixteen sub-matrices each has a 2τ bandwidth.

The $\lambda_2{\lambda_4}^{-1}$ consists of four sub-matrices of limited 2τ bandwidth with $N(4096\tau^2 + 512\tau + 64)$ flops. In the similar way, $\lambda_2{\lambda_4}^{-1}\lambda_3$ building necessitates roughly $N(4096\tau^2 + 512\tau + 64)$ flops of limited 2τ bandwidth for each of the sixteen sub-matrices. The matrix subtraction necessitates $N(64\tau + 8)$ flops.

Now, matrix inversion is used to create Φ with a bandwidth of 2τ. This is equivalent to $N(4688\tau^2 + 1284\tau + 152)$ flops. As a result, the Φ matrix construction requires about $N(17568\tau^2 + 3656\tau + 440)$ flops.

*Construction of $\bar{\lambda}_2$ matrix*

The $\bar{\lambda}_2$ construction needs a complex matrix by matrix multiplication of the $\Phi$ matrix, and the constructed matrix later $\left(\text{i.e., } \lambda_2{\lambda_4}^{-1}\right)$, moreover each of them $\left(\text{i.e., } \lambda_2{\lambda_4}^{-1},\ \Phi\right)$ consists of sixteen sub-matrices. This equates to $N(1024\tau^2 + 512\tau + 64)$ flops of limited $2\tau$ bandwidth.

<u>*Construction of $\bar{\lambda}_3$ matrix*</u>

The $\bar{\lambda}_3$ construction needs complex matrix by matrix multiplication of ${\lambda_4}^{-1}\lambda_3$ and $\Phi$. Firstly, the construction of the ${\lambda_4}^{-1}\lambda_3$ matrix needs $N(1024\tau^2 + 512\tau + 64)$ flops of limited $2\tau$ bandwidth. This is followed by the multiplication of the $\Phi$ matrix, which needs $N(1024\tau^2 + 512\tau + 64)$ flops of limited $2\tau$ bandwidth. Thus, the construction of the $\bar{\lambda}_3$ matrix needs $N(2048\tau^2 + 1024\tau + 128)$ flops of limited to $2\tau$ bandwidth.

<u>*Construction of $\bar{\lambda}_4$ matrix*</u>

In fact, $\bar{\lambda}_4 = {\lambda_4}^{-1} + \bar{\lambda}_3\, \lambda_2\, {\lambda_4}^{-1}$ , which equates to $N(1024\tau^2 + 576\tau + 72)$ flops. The creation of $\bar{\lambda}_i$ matrices require roughly $N(21664\tau^2 + 5768\tau + 704)$ flops in total. Meanwhile, the construction of the $(\boldsymbol{\mu}^H\boldsymbol{\mu} + \alpha\, \mathbf{I}_{4N\times 4N})$ matrix requires $2^{2\sigma}N(16\tau^2 + 8\tau + 4)$+$2^{\sigma}N$, flops $\sigma = 3$.

The generated matrix$(\text{i.e., } (\boldsymbol{\mu}^H\boldsymbol{\mu} + \alpha\, \mathbf{I}_{8N\times 8N})^{-1})$ of $2\tau$ limited bandwidth is multiplied by the $\boldsymbol{\mu}^H$ matrix to get the JLCRLZF-CDE solution matrix stated in Eq. (19). This is equivalent to $2^{2\sigma}N(144\tau^2 + 24\tau + 4)$ flops, with $3\tau$ bandwidth, which is limited also to $2\tau$ bandwidth. As a result, the JLCRLZF-CDE solution matrix requires roughly $N(31904\tau^2 + 7816\tau + 1224)$ flops to create. The receiving vector is multiplied by the JLCRLZF-CDE solution matrix, which requires $2^{2\sigma}N(16\tau + 8)$ flops. As a result, the JLCRLZF-CDE necessitates $N(31904\tau^2 + 8840\tau + 1736)$ flops. The JLCRLZF-CDE takes 8.491 m min to complete its operation on each run, according to the simulation parameters in Table 1 and using a laptop with a core i5 CPU.

*iii.* 16×16 *MIMO-DCT-OFDM system*

The matrix inversion in the case of 16×16 MIMO-DCT-OFDM system consists of 256 sub-matrices $(\text{i.e., } 2^4 \times 2^4 = 256)$, which can be written as:

$$(\boldsymbol{\mu}^H\boldsymbol{\mu} + \alpha\,\mathbf{I}_{16N\times 16N})^{-1} = \begin{bmatrix} \overbrace{\begin{bmatrix} \lambda_{1,1} & \lambda_{1,2} & \cdots & \lambda_{1,8} \\ \lambda_{2,1} & \lambda_{2,2} & \cdots & \lambda_{2,8} \\ \vdots & \vdots & \ddots & \vdots \\ \lambda_{8,1} & \lambda_{8,2} & \cdots & \lambda_{8,8} \end{bmatrix}}^{\lambda_1} & \overbrace{\begin{bmatrix} \lambda_{1,9} & \lambda_{1,10} & \cdots & \lambda_{1,16} \\ \lambda_{2,9} & \lambda_{2,10} & \cdots & \lambda_{2,16} \\ \vdots & \vdots & \ddots & \vdots \\ \lambda_{8,9} & \lambda_{8,10} & \cdots & \lambda_{8,16} \end{bmatrix}}^{\lambda_2} \\ \underbrace{\begin{bmatrix} \lambda_{9,1} & \lambda_{9,2} & \cdots & \lambda_{9,8} \\ \lambda_{10,1} & \lambda_{10,2} & \cdots & \lambda_{10,8} \\ \vdots & \vdots & \ddots & \vdots \\ \lambda_{16,1} & \lambda_{16,2} & \cdots & \lambda_{16,8} \end{bmatrix}}_{\lambda_3} & \underbrace{\begin{bmatrix} \lambda_{9,9} & \lambda_{9,10} & \cdots & \lambda_{9,16} \\ \lambda_{10,9} & \lambda_{10,10} & \cdots & \lambda_{10,16} \\ \vdots & \vdots & \ddots & \vdots \\ \lambda_{16,9} & \lambda_{16,10} & \cdots & \lambda_{16,16} \end{bmatrix}}_{\lambda_4} \end{bmatrix}^{-1} \quad (26)$$

$$= \begin{bmatrix} \bar{\lambda}_1 & \bar{\lambda}_2 \\ \bar{\lambda}_3 & \bar{\lambda}_4 \end{bmatrix} = \begin{bmatrix} \Phi & -\Phi\lambda_2{\lambda_4}^{-1} \\ -{\lambda_4}^{-1}\lambda_3\Phi & {\lambda_4}^{-1} + {\lambda_4}^{-1}\lambda_3\Phi\lambda_2{\lambda_4}^{-1} \end{bmatrix}$$

*Construction of $\bar{\lambda}_1$ matrix*

$\bar{\lambda}_1 = \Phi = \left(\lambda_1 - \lambda_2{\lambda_4}^{-1}\lambda_3\right)^{-1}, \lambda_i, i \in \{1, 2, 3, 4\}$ has $16N \times 16N$ dimensions, which consists of 256 sub-matrices each with a bandwidth of 2τ.

Let's figure out how many flops are required for the $\bar{\lambda}_i, i \in \{1,\ 2, 3,\ 4\}$ structure. In fact, the ${\lambda_4}^{-1}$ consists of 256 sub-matrices each with bandwidth of 2τ, which can construct using the same steps of the 2×2, 4×4, and 8×8 system configuration described in Eqs. (23), (24), and (25), respectively. It is should note that, each constructed matrix exceeds 2τ bandwidth is limited to 2τ bandwidth. Thus, the ${\lambda_4}^{-1}$ matrix construction needs $N(21664\tau^2 + 5768\tau + 704)$ flops, which consists of 256 sub-matrices each has a 2τ bandwidth.

The $\lambda_2{\lambda_4}^{-1}$ consists of four sub-matrices of limited 2τ bandwidth with $N(16384\tau^2 + 2048\tau + 256)$ flops. In the similar way, $\lambda_2{\lambda_4}^{-1}\lambda_3$ building necessitates roughly $N(16384\tau^2 + 2048\tau + 256)$ flops of limited 2τ bandwidth for each of the 256 sub-matrices. The matrix subtraction necessitates $N(256\tau + 32)$ flops.

Now, matrix inversion is used to create Φ with a bandwidth of 2τ. This is equivalent to $N(21664\tau^2 + 5768\tau + 704)$ flops. As a result, the Φ matrix construction requires about $N(76096\tau^2 + 15888\tau + 1952)$ flops.

*Construction of $\bar{\lambda}_2$ matrix*

The $\bar{\lambda}_2$ construction needs a complex matrix by matrix multiplication of the $\Phi$ matrix, and the constructed matrix later $\left(\text{i. e.,}\ \lambda_2\lambda_4{}^{-1}\right)$, moreover each of them $\left(\text{i. e.,}\ \lambda_2\lambda_4{}^{-1},\ \Phi\right)$ consists of 64 sub-matrices. This equates to $N(4096\tau^2 + 2048\tau + 256)$ flops of limited $2\tau$ bandwidth.

*Construction of $\bar{\lambda}_3$ matrix*

The $\bar{\lambda}_3$ construction needs complex matrix by matrix multiplication of $\lambda_4{}^{-1}\lambda_3$ and $\Phi$. Firstly, the construction of the $\lambda_4{}^{-1}\lambda_3$ matrix needs $N(4096\tau^2 + 2048\tau + 256)$ flops of limited $2\tau$ bandwidth. This is followed by the multiplication of the $\Phi$ matrix, which needs $N(4096\tau^2 + 2048\tau + 256)$ flops of limited $2\tau$ bandwidth. Thus, the construction of the $\bar{\lambda}_3$ matrix needs $N(8192\tau^2 + 4096\tau + 512)$ flops of limited to $2\tau$ bandwidth.

*Construction of $\bar{\lambda}_4$ matrix*

In fact, $\bar{\lambda}_4 = \lambda_4{}^{-1} + \bar{\lambda}_3\ \lambda_2\ \lambda_4{}^{-1}$, which equates to $N(4096\tau^2 + 2304\tau + 288)$ flops. The creation of $\bar{\lambda}_i$ matrices require roughly $N(92480\tau^2 + 24336\tau + 3008)$ flops in total. Call that, the construction of the $(\boldsymbol{\mu}^H\boldsymbol{\mu} + \alpha\,\mathbf{I}_{4N\times 4N})$ matrix requires $2^{2\sigma}N(16\tau^2 + 8\tau + 4)$+$2^{\sigma}N$ flops, $\sigma = 4$. The generated matrix$(\text{i. e.,}\ (\boldsymbol{\mu}^H\boldsymbol{\mu} + \alpha\,\mathbf{I}_{16N\times 16N})^{-1})$ of $2\tau$ limited bandwidth is multiplied by the $\boldsymbol{\mu}^H$ matrix to get the JLCRLZF-CDE solution matrix stated in Eq. (19). This is equivalent to $2^{2\sigma}N(144\tau^2 + 24\tau + 4)$ flops, with $3\tau$ bandwidth, which is limited also to $2\tau$ bandwidth. As a result, the JLCRLZF-CDE solution matrix requires roughly $N(133440\tau^2 + 32528\tau + 5072)$ flops to create. The receiving vector is multiplied by the JLCRLZF-CDE solution matrix, which requires $2^{2\sigma}N(16\tau + 8)$ flops. As a result, the JLCRLZF-CDE necessitates $N(133440\tau^2 + 36624\tau + 7120)$ flops. The JLCRLZF-CDE takes 42.490 m min to complete its operation on each run, according to the simulation parameters in Table 1 and using a laptop with a core i5 CPU.

*iv.* 32×32 *MIMO-DCT-OFDM system*

The matrix inversion in the case of 32×32 MIMO-DCT-OFDM system consists of 1024 sub-matrices $(\text{i. e.,}\ 2^5 \times 2^5 = 1024)$, which can be written as:

$$
(\boldsymbol{\mu}^H \boldsymbol{\mu} + \alpha\, \mathbf{I}_{32N\times 32N})^{-1} = \left[\begin{array}{cc} \overbrace{\begin{array}{|cccc|} \hline \lambda_{1,1} & \lambda_{1,2} & \cdots & \lambda_{1,16} \\ \lambda_{2,1} & \lambda_{2,2} & \cdots & \lambda_{2,16} \\ \vdots & \vdots & \ddots & \vdots \\ \lambda_{16,1} & \lambda_{16,2} & \cdots & \lambda_{16,16} \\ \hline \end{array}}^{\lambda_1} & \overbrace{\begin{array}{|cccc|} \hline \lambda_{1,17} & \lambda_{1,18} & \cdots & \lambda_{1,32} \\ \lambda_{2,17} & \lambda_{2,18} & \cdots & \lambda_{2,32} \\ \vdots & \vdots & \ddots & \vdots \\ \lambda_{16,17} & \lambda_{16,18} & \cdots & \lambda_{16,32} \\ \hline \end{array}}^{\lambda_2} \\ \underbrace{\begin{array}{|cccc|} \hline \lambda_{17,1} & \lambda_{17,2} & \cdots & \lambda_{17,16} \\ \lambda_{18,1} & \lambda_{18,2} & \cdots & \lambda_{18,16} \\ \vdots & \vdots & \ddots & \vdots \\ \lambda_{32,1} & \lambda_{32,2} & \cdots & \lambda_{32,16} \\ \hline \end{array}}_{\lambda_3} & \underbrace{\begin{array}{|cccc|} \hline \lambda_{17,17} & \lambda_{17,18} & \cdots & \lambda_{17,32} \\ \lambda_{18,17} & \lambda_{18,18} & \cdots & \lambda_{18,32} \\ \vdots & \vdots & \ddots & \vdots \\ \lambda_{32,17} & \lambda_{32,18} & \cdots & \lambda_{32,32} \\ \hline \end{array}}_{\lambda_4} \end{array}\right]^{-1} \tag{27}
$$

$$
= \begin{bmatrix} \bar{\lambda}_1 & \bar{\lambda}_2 \\ \bar{\lambda}_3 & \bar{\lambda}_4 \end{bmatrix} = \begin{bmatrix} \Phi & -\Phi \lambda_2 {\lambda_4}^{-1} \\ -{\lambda_4}^{-1} \lambda_3 \Phi & {\lambda_4}^{-1} + {\lambda_4}^{-1} \lambda_3 \Phi \lambda_2 {\lambda_4}^{-1} \end{bmatrix}
$$

*Construction of $\bar{\lambda}_1$ matrix*

Using the same procedure as later. Thus, the $\Phi$ matrix construction requires about $N(316032\tau^2 + 66080\tau + 8192)$ flops.

*Construction of $\bar{\lambda}_2$ matrix*

The $\bar{\lambda}_2$ construction consists of 256 sub-matrices. This equates to $N(16384\tau^2 + 8192\tau + 1024)$ flops of limited $2\tau$ bandwidth.

*Construction of $\bar{\lambda}_3$ matrix*

The $\bar{\lambda}_3$ construction needs $N(32768\tau^2 + 16384\tau + 2048)$ flops of limited to $2\tau$ bandwidth.

*Construction of $\bar{\lambda}_4$ matrix*

The $\bar{\lambda}_4$ matrix construction needs about $N(16384\tau^2 + 9216\tau + 1152)$ flops. Thus, the creation of $\bar{\lambda}_i$ matrices require roughly $N(381568\tau^2 + 99872\tau + 12416)$ flops in total. The construction of the $(\boldsymbol{\mu}^H \boldsymbol{\mu} + \alpha\, \mathbf{I}_{4N\times 4N})$ matrix requires $2^{2\sigma} N(16\tau^2 + 8\tau + 4)$+$2^{\sigma} N$ flops, $\sigma = 5$.

The generated matrix(i.e., $(\boldsymbol{\mu}^H \boldsymbol{\mu} + \alpha\, \mathbf{I}_{32N\times 32N})^{-1}$) of $2\tau$ limited bandwidth is multiplied by the $\boldsymbol{\mu}^H$ matrix to get the JLCRLZF-CDE solution matrix stated in Eq. (19). This is equivalent to $2^{2\sigma} N(144\tau^2 + 24\tau + 4)$ flops, with $3\tau$ bandwidth, which is limited also to $2\tau$ bandwidth. As a result, the JLCRLZF-CDE solution matrix requires roughly $N(545408\tau^2 + 132640\tau + 20640)$ flops to create. The receiving vector is multiplied by the JLCRLZF-CDE solution matrix, which requires $2^{2\sigma}N(16\tau+8)$ flops. As a result, the JLCRLZF-CDE necessitates $N(545408\tau^2+149024\tau+28832)$ flops. The JLCRLZF-CDE takes 173.407 m min to complete its operation on each run, according to the simulation parameters in Table 1 and using a laptop with a core i5 CPU.

*v.* 64×64 *MIMO-DCT-OFDM system*

The matrix inversion in the case of 64×64 MIMO-DCT-OFDM system consists of 4096 sub-matrices ($\text{i.e.,}\ 2^6\times 2^6=4096$), which can be written as:

$$(\boldsymbol{\mu}^H\boldsymbol{\mu}+\alpha\,\mathbf{I}_{64N\times 64N})^{-1}=\left[\begin{array}{cc}\overbrace{\begin{bmatrix}\lambda_{1,1} & \lambda_{1,2} & \cdots & \lambda_{1,32}\\ \lambda_{2,1} & \lambda_{2,2} & \cdots & \lambda_{2,32}\\ \vdots & \vdots & \ddots & \vdots\\ \lambda_{32,1} & \lambda_{32,2} & \cdots & \lambda_{32,32}\end{bmatrix}}^{\lambda_1} & \overbrace{\begin{bmatrix}\lambda_{1,33} & \lambda_{1,34} & \cdots & \lambda_{1,64}\\ \lambda_{2,33} & \lambda_{2,34} & \cdots & \lambda_{2,64}\\ \vdots & \vdots & \ddots & \vdots\\ \lambda_{32,33} & \lambda_{32,34} & \cdots & \lambda_{32,64}\end{bmatrix}}^{\lambda_2}\\ \underbrace{\begin{bmatrix}\lambda_{33,1} & \lambda_{33,2} & \cdots & \lambda_{33,32}\\ \lambda_{34,1} & \lambda_{34,2} & \cdots & \lambda_{34,32}\\ \vdots & \vdots & \ddots & \vdots\\ \lambda_{64,1} & \lambda_{64,2} & \cdots & \lambda_{64,32}\end{bmatrix}}_{\lambda_3} & \underbrace{\begin{bmatrix}\lambda_{33,33} & \lambda_{33,34} & \cdots & \lambda_{33,64}\\ \lambda_{34,33} & \lambda_{34,34} & \cdots & \lambda_{34,64}\\ \vdots & \vdots & \ddots & \vdots\\ \lambda_{64,33} & \lambda_{64,34} & \cdots & \lambda_{64,64}\end{bmatrix}}_{\lambda_4}\end{array}\right]^{-1} \quad (28)$$

$$=\begin{bmatrix}\bar{\lambda}_1 & \bar{\lambda}_2\\ \bar{\lambda}_3 & \bar{\lambda}_4\end{bmatrix}=\begin{bmatrix}\Phi & -\Phi\lambda_2{\lambda_4}^{-1}\\ -{\lambda_4}^{-1}\lambda_3\Phi & {\lambda_4}^{-1}+{\lambda_4}^{-1}\lambda_3\Phi\lambda_2{\lambda_4}^{-1}\end{bmatrix}$$

Using the same procedure as later. The $\bar{\lambda}_1$, and $\bar{\lambda}_2$ matrices construction require about $N(1287424\tau^2+269376\tau+33536)$, and $N(65536\tau^2+32768\tau+4096)$ flops, repectively of limited 2τ bandwidth for each.

In the same manner, the $\bar{\lambda}_3$, and $\bar{\lambda}_4$ matrices construction need $N(131072\tau^2+65536\tau+8192)$, and $N(65536\tau^2+36864\tau+4608)$ flops, respectively of limited 2τ bandwidth for each. Thus, the creation of $\bar{\lambda}_i$ matrices require roughly $N(1549568\tau^2+404544\tau+50432)$ flops in total.

The JLCRLZF-CDE solution matrix stated in Eq. (19) in the case of 64×64 MIMO-DCT-OFDM system requires roughly $N(2204928\tau^2+601152\tau+116032)$ flops. The JLCRLZF-CDE takes 762.613 m min to complete its operation on each run, according to the simulation parameters in Table 1 and using a laptop with a core i5 CPU.

*vi.* $2^\sigma\times 2^\sigma$ *MIMO-DCT-OFDM system*

The matrix inversion in the case of $2^{\sigma} \times 2^{\sigma}$ MIMO-DCT-OFDM system consists of $2^{2\sigma}$ sub-matrices, which can be written as:

$$(\boldsymbol{\mu}^H \boldsymbol{\mu} + \alpha\, \mathbf{I}_{2^{\sigma}N \times 2^{\sigma}N})^{-1}$$

$$= \left[\begin{array}{cc} \overbrace{\begin{bmatrix} \lambda_{1,1} & \lambda_{1,2} & \cdots & \lambda_{1,2^{\sigma-1}} \\ \lambda_{2,1} & \lambda_{2,2} & \cdots & \lambda_{2,2^{\sigma-1}} \\ \vdots & \vdots & \ddots & \vdots \\ \lambda_{2^{\sigma-1},1} & \lambda_{2^{\sigma-1},2} & \cdots & \lambda_{2^{\sigma-1},2^{\sigma-1}} \end{bmatrix}}^{\lambda_1} & \overbrace{\begin{bmatrix} \lambda_{1,2^{\sigma-1}+1} & \lambda_{1,2^{\sigma-1}+2} & \cdots & \lambda_{1,2^{\sigma}} \\ \lambda_{2,2^{\sigma-1}+1} & \lambda_{2,2^{\sigma-1}+2} & \cdots & \lambda_{2,2^{\sigma}} \\ \vdots & \vdots & \ddots & \vdots \\ \lambda_{2^{\sigma-1},2^{\sigma-1}+1} & \lambda_{2^{\sigma-1},2^{\sigma-1}+2} & \cdots & \lambda_{2^{\sigma-1},2^{\sigma}} \end{bmatrix}}^{\lambda_2} \\ \underbrace{\begin{bmatrix} \lambda_{2^{\sigma-1}+1,1} & \lambda_{2^{\sigma-1}+1,2} & \cdots & \lambda_{2^{\sigma-1}+1,2^{\sigma-1}} \\ \lambda_{2^{\sigma-1}+2,1} & \lambda_{2^{\sigma-1}+2,2} & \cdots & \lambda_{2^{\sigma-1}+2,2^{\sigma-1}} \\ \vdots & \vdots & \ddots & \vdots \\ \lambda_{2^{\sigma},1} & \lambda_{2^{\sigma},2} & \cdots & \lambda_{2^{\sigma},2^{\sigma-1}} \end{bmatrix}}_{\lambda_3} & \underbrace{\begin{bmatrix} \lambda_{2^{\sigma-1}+1,2^{\sigma-1}+1} & \lambda_{2^{\sigma-1}+1,2^{\sigma-1}+2} & \cdots & \lambda_{2^{\sigma-1}+1,2^{\sigma}} \\ \lambda_{2^{\sigma-1}+2,2^{\sigma-1}+1} & \lambda_{2^{\sigma-1}+2,2^{\sigma-1}+2} & \cdots & \lambda_{2^{\sigma-1}+2,2^{\sigma}} \\ \vdots & \vdots & \ddots & \vdots \\ \lambda_{2^{\sigma},2^{\sigma-1}+1} & \lambda_{2^{\sigma},2^{\sigma-1}+2} & \cdots & \lambda_{2^{\sigma},2^{\sigma}} \end{bmatrix}}_{\lambda_4} \end{array}\right]^{-1} \quad (29)$$

$$= \begin{bmatrix} \bar{\lambda}_1 & \bar{\lambda}_2 \\ \bar{\lambda}_3 & \bar{\lambda}_4 \end{bmatrix} = \begin{bmatrix} \Phi & -\Phi\lambda_2{\lambda_4}^{-1} \\ -{\lambda_4}^{-1}\lambda_3\Phi & {\lambda_4}^{-1} + {\lambda_4}^{-1}\lambda_3\Phi\lambda_2{\lambda_4}^{-1} \end{bmatrix}$$

*Construction of $\bar{\lambda}_1$ matrix*

Using the same procedure as later. The $(\Phi)_{2^{\sigma}\times 2^{\sigma}}$ matrix construction requires the same number of flops needed for the $(\bar{\lambda}_i)_{2^{\sigma-1}\times 2^{\sigma-1}}$ construction.

*Construction of $\bar{\lambda}_2$ matrix*

The $(\bar{\lambda}_2)_{2^{\sigma}\times 2^{\sigma}}$ construction consists of $2^{2\sigma-2}$ sub-matrices, which requires the same number of flops needed for the $\left(4 \times (\bar{\lambda}_2)_{2^{\sigma-1}\times 2^{\sigma-1}}\right)$ construction.

*Construction of $\bar{\lambda}_3$ matrix*

The $(\bar{\lambda}_3)_{2^{\sigma}\times 2^{\sigma}}$ construction consists of $2^{2\sigma-2}$ sub-matrices, which requires the same number of flops needed for the $\left(4 \times (\bar{\lambda}_3)_{2^{\sigma-1}\times 2^{\sigma-1}}\right)$ construction.

*Construction of $\bar{\lambda}_4$ matrix*

The $(\bar{\lambda}_4)_{2^{\sigma}\times 2^{\sigma}}$ construction consists of $2^{2\sigma-2}$ sub-matrices, which requires the same number of flops needed for the $\left(4 \times (\bar{\lambda}_4)_{2^{\sigma-1}\times 2^{\sigma-1}}\right)$ construction. Note that, all sub-matrices bandlimited to $2\tau$ as explained in the lower order configurations.

Thus, the creation of $(\bar{\lambda}_i)_{2^\sigma \times 2^\sigma}$ matrices require roughly $\sum_{q=1}^{4} (\bar{\lambda}_q)_{2^\sigma \times 2^\sigma}$ flops in total.

The generated matrix $(\text{i.e.,}\ (\boldsymbol{\mu}^H \boldsymbol{\mu} + \alpha\, \mathbf{I}_{2^\sigma N \times 2^\sigma N})^{-1})$ of $2\tau$ limited bandwidth is multiplied by the $\boldsymbol{\mu}^H$ matrix to get the JLCRLZF-CDE solution matrix stated in Eq. (19). This is equivalent to $2^{2\sigma} N(144\tau^2 + 24\tau + 4)$ flops, with $3\tau$ bandwidth, which is limited also to $2\tau$ bandwidth. The receiving vector is multiplied by the JLCRLZF-CDE solution matrix, which requires $2^{2\sigma} N(16\tau + 8)$ flops.

**Table 2**. The number of flops corresponds to different mathematical operations [21].

| | Number of | | | | |
|---|---|---|---|---|---|
| Process | × | ∓ | ÷ | Operations | Flops |
| $a + b$ | 0 | 1 | 0 | 1 | 0.5 |
| $a - b$ | 0 | 1 | 0 | 1 | 0.5 |
| $a.b$ | 1 | 0 | 0 | 1 | 0.5 |
| $a/b$ | 0 | 0 | 1 | 1 | 0.5 |
| $a + (b + jc)$ | 1 | 1 | 0 | 1 | 0.5 |
| $a.(b + jc)$ | 2 | 0 | 0 | 2 | 1 |
| $(a + jb) + (c + jd)$ | 0 | 2 | 0 | 2 | 1 |
| $(a + jb).(c + jd)$ | 4 | 2 | 0 | 6 | 3 |

**Table 3**. The number of flops corresponds to different full-matrix operations [22].

| Process | Description | Full-matrix computation | | | | | | |
|---|---|---|---|---|---|---|---|---|
| | | Number of complex | | Number of real-complex | | | Total number of | |
| | | × | ∓ | × | ∓ | ÷ | Operations | Flops |
| $\boldsymbol{\mathcal{A}} \mp \boldsymbol{\mathcal{B}}$ | $\boldsymbol{\mathcal{A}}, \boldsymbol{\mathcal{B}} \in \mathbb{C}^{N\times N}$ | --- | $2N^2$ | --- | --- | --- | $2N^2$ | $N^2$ |
| $\boldsymbol{\mathcal{A}}.\boldsymbol{\mathcal{B}}$ | | $N^3$ | $N^3$ | --- | --- | --- | $8N^3$ | $4N^3$ |
| $\boldsymbol{\mathcal{A}}^{-1}$ | | $N^3$ | $N^3$ | $N^2$ | --- | $N$ | $8N^3 + 2N^2 + N$ | $4N^3 + N^2 + N/2$ |
| $\boldsymbol{\mathcal{A}}.\boldsymbol{\mathcal{U}}$ | $\boldsymbol{\mathcal{U}} \in \mathbb{C}^{N\times 1}$ | $N^2$ | $N^2$ | --- | --- | --- | $8N^2$ | $4N^2$ |

**Table 4**. The number of flops corresponds to different banded-matrix operations.

| Process | Description | Banded-matrix computation | | | | | | | |
|---|---|---|---|---|---|---|---|---|---|
| | | Number of real | Number of complex | | | Number of real-complex | | Total number of | |
| | | × | × | ∓ | ÷ | × | ∓ | Operations | Flops |
| $\boldsymbol{\mathcal{A}} \mp \boldsymbol{\mathcal{B}}$ | $\boldsymbol{\mathcal{A}}, \boldsymbol{\mathcal{B}} \in \mathbb{C}^{N\times N}$ | -- | -- | $N(4\tau+1)$ | -- | -- | -- | $N(4\tau+1)$ | $N(2\tau+0.5)$ |
| $\boldsymbol{\mathcal{A}}.\boldsymbol{\mathcal{B}}$ | | -- | $N(2\tau+1)^2$ | $N(2\tau+1)^2$ | -- | -- | -- | $N(32\tau^2+16\tau+8)$ | $N(16\tau^2+8\tau+4)$ |
| $\boldsymbol{\mathcal{A}}^{-1}$ | | -- | $N(\tau^2+2\tau)$ | $N(\tau^2+2\tau)$ | $N(\tau+1)$ | -- | -- | $N(10\tau^2+21\tau+1)$ | $N(5\tau^2+10.5\tau+0.5)$ |
| $\boldsymbol{\mathcal{A}}.\boldsymbol{\mathcal{U}}$ | $\boldsymbol{\mathcal{U}} \in \mathbb{C}^{N\times 1}$ | -- | $N(12\tau+6)$ | $N(4\tau+2)$ | -- | -- | -- | $N(16\tau+8)$ | $N(8\tau+4)$ |

Table 5 gives the number of flops for different equalizers configurations of $2^\sigma \times 2^\sigma$ MIMO-DCT-OFDM system. Table 6 gives the average simulated time of various equalization procedures for different configurations of $2^\sigma \times 2^\sigma$ MIMO-DCT-OFDM system. Also, the simulated time reduction percentage of each equalizer type of $2^{\sigma} \times 2^{\sigma}$ MIMO-DCT-OFDM system is tabulated in Table 7 with respect to that of proposed JLCRLZF-CDE, which can be expressed as:

$$\eta\% = \frac{t_2 - t_1}{t_1}\% \tag{24}$$

where $t_1$ is the simulation time of the proposed JLCRLZF-CDE, and $t_2$ is the simulation of the compared equalizer. According to Table 7, it is clear that the proposed JLCRLZF-CDE saves about 2.12%, 74.11%, 96.64%, 93.04%, 98.86%, 140.21%, and 158.03% of the simulated time compared to that of LZF-CDE, for that of the configuration cases 2×2, 4×4, 8×8, 16×16, 32×32, 64×64, and 128×128, respectively. In the same manner, the proposed JLCRLZF-CDE saves about 20.30%, 83.73%, 101.93%, 109.41%, 142.49%, 159.18%, 191.90% of the simulated time compared to that of LMMSE-FDE, for that of the configuration cases 2×2, 4×4, 8×8, 16×16, 32×32, 64×64, and 128×128, respectively. Also, the simulated reduction time of other schemes compared to that of the proposed JLCRLZF-CDE is tabulated in Table 7.

**Table 5**. The number of flops, and the average simulated time of various equalization procedures for different configurations of a $2^{\sigma} \times 2^{\sigma}$ MIMO-DCT-OFDM system.

| Equalizer type | Number of flops | SNR estimation |
|---|---|---|
| LZF-CDE | $12M^3 + 5M^2 + M/2$ | × |
| LMMSE-CDE | $12M^3 + 5M^2 + 2.5M$ | ✓ |
| LZF-FDE | $12M^3 + 6M^2 - M/2$ | × |
| LMMSE-FDE | $12M^3 + 6M^2 + 2M$ | ✓ |
| JLCRLZF-CDE | $\sum_{q=1}^{4} (\bar{\lambda}_q)_{2^{\sigma}\times 2^{\sigma}} + 2^{2\sigma}N(144\tau^2 + 40\tau + 12)$ | × |

**Table 6**. The average simulated time of various equalization procedures for different configurations of a $2^{\sigma} \times 2^{\sigma}$ MIMO-DCT-OFDM system.

| Equalizer type | Simulation time for each run (m min) | | | | | | |
|---|---|---|---|---|---|---|---|
| | $\sigma = 1$ | $\sigma = 2$ | $\sigma = 3$ | $\sigma = 4$ | $\sigma = 5$ | $\sigma = 6$ | $\sigma = 7$ |
| LZF-CDE | 0.337 | 4.142 | 16.442 | 76.323 | 344.851 | 1831.907 | 12570.318 |
| LMMSE-CDE | 0.365 | 4.223 | 16.512 | 74.806 | 386.557 | 1876.315 | 12888.468 |
| LZF-FDE | 0.357 | 4.311 | 16.820 | 78.084 | 415.417 | 1958.331 | 13870.996 |
| LMMSE-FDE | 0.397 | 4.371 | 17.146 | 82.794 | 420.496 | 1976.542 | 14220.691 |
| JLCRLZF-CDE | 0.330 | 2.379 | 8.491 | 42.490 | 173.407 | 762.613 | 4871.709 |

**Table 7**. The time reduction percentage for different configurations of a $2^{\sigma} \times 2^{\sigma}$ MIMO-DCT-OFDM system with respect to the proposed JLCRLZF-CDE.

| Equalizer type | η % | | | | | | |
|---|---|---|---|---|---|---|---|
| | $\sigma = 1$ | $\sigma = 2$ | $\sigma = 3$ | $\sigma = 4$ | $\sigma = 5$ | $\sigma = 6$ | $\sigma = 7$ |
| LZF-CDE | 2.12 | 74.11 | 96.64 | 93.04 | 98.86 | 140.21 | 158.03 |
| LMMSE-CDE | 10.61 | 77.51 | 94.46 | 94.84 | 122.92 | 146.04 | 164.56 |
| LZF-FDE | 8.18 | 81.21 | 98.09 | 97.50 | 139.56 | 156.79 | 184.73 |
| LMMSE-FDE | 20.30 | 83.73 | 101.93 | 109.41 | 142.49 | 159.18 | 191.90 |
| JLCRLZF-CDE | ---- | ---- | ---- | ---- | ---- | ---- | ---- |

## VI. Conclusions

In this work, we proposed a JLCRLZF-CDE for a $2^{\sigma} \times 2^{\sigma}$ MIMO-DCT-OFDM system and compared it to various equalizers in a variety of ways, $\sigma \in \{1,2,3,4, \dots \dots\}$. Using the banded-matrix approximation, the proposed JLCRLZF-CDE simultaneously executes the equalization and CFO compensation algorithms with decreased complexity. The proposed JLCRLZF-CDE was designed with the co-CFO, ISI, co-channel interference, and noise in mind. Furthermore, the advantages of the proposed JLCRLZF-CDE lies in its simplicity besides the computer simulations have been performed, and some situations have been discovered in which the proposed JLCRLZF-CDE is more effective in terms of average simulation time than other equalizers. The main challenge of the proposed JLCRLZF-CDE is that the lower performance than that of the LMMSE-FDE within the variation range of the SNR (i.e., SNR=0:5:25 dB). This appears to be a compromise between complexity and BER performance.

## VII. Acknowledgement

My deepest love and gratitude are devoted to my whole family. I would like to thank my father, Ramadan, my mother, my wife, my brother Mohamed, my sisters, and my sons, Omar and Retal, for their infinite patience and trust. They were usually beside me in all the happy as well as hard times.

## VIII. conflicts of interest

Not applicable

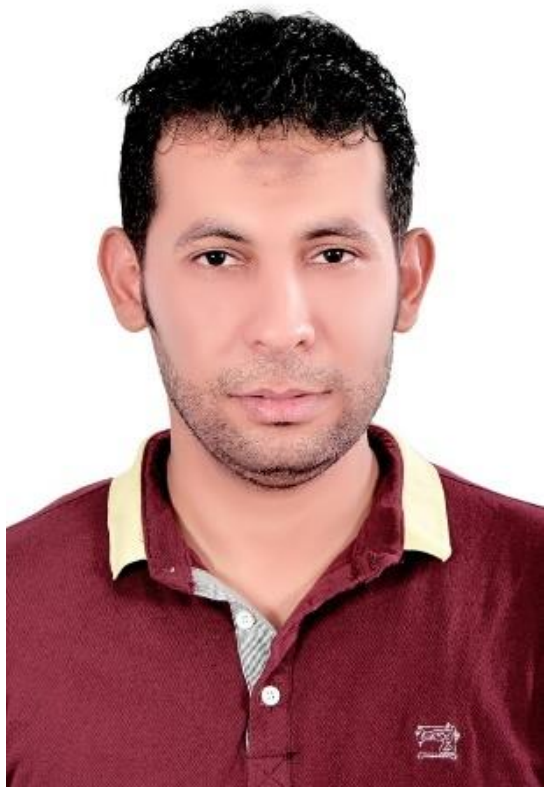

**Khaled Ramadan Mohamed** received his B.Sc. from the Higher Institute of Engineering (El-Shorouk Academy) in 2011, and his M.Sc. and Ph.D. from the Faculty of Electronic Engineering, Menoufia University, Menouf, Egypt, in 2018 and 2021, respectively. He joined the teaching staff of the Department of Electronics and Electrical Communications at the Higher Institute of Engineering in Al-Shorouk City, Egypt, in 2021. He has published several scientific papers in national and international conference proceedings and journals. He has received the most read paper from the International Journal of Communication Systems (IJCS) for 2018–2019. His research interests include multicarrier communication systems, Multiple-Input Multiple-Output (MIMO) systems, Digital Signal Processing (DSP), digital communications, channel equalization, Carrier Frequency Offsets (CFOs) estimations and compensations, Underwater Acoustic (UWA) wireless communication systems, and multicarrier Non-Orthogonal Multiple Access (NOMA) for 5G networks.